\documentclass[aps,prx,twocolumn,groupedaddress,amsmath,epsfig,amssymb,eqsecnum]{revtex4}
\usepackage{makecell}
\usepackage{bbm}
\usepackage{mathrsfs}
\usepackage{subfigure}
\usepackage{enumerate}
\usepackage{amsfonts}
\usepackage{color}
\usepackage{graphicx}
\newcommand{\xing}[1] {{\color{black} #1}}

\newcommand{\dbar}{d\hspace*{-0.08em}\bar{}\hspace*{0.1em}}

\newcommand{\Xv}{\mathbf X}

\newcommand{\rv}{{\mathbf r}}

\newcommand{\xv}{\boldsymbol x}

\newcommand{\qv}{{\boldsymbol q}}
\newcommand{\pv}{{\boldsymbol p}}
\newcommand{\yv}{{\boldsymbol y}}

\newcommand{\Qm}{{\boldsymbol{Q}}}

\newcommand{\Lm}{{\boldsymbol L}}

\newcommand{\Bm}{{\boldsymbol B}}

\newcommand{\be}{\begin{equation}}
\newcommand{\ee}{\end{equation}}
\newcommand{\ba}{\begin{eqnarray}}
\newcommand{\ea}{\end{eqnarray}}
\newcommand{\pdf}{\mathcal{p}}
\newcommand{\U}{\mathcal{U}}
\newcommand{\LFP}{\mathcal{L}_{\rm FP}}

    \usepackage{boondox-cal}
\begin{document}
\title{Covariant Non-equilibrium Thermodynamics from Ito-Langevin Dynamics}
\author{Mingnan Ding$^{1}$}
\author{Xiangjun Xing$^{1,3,4}$}
\email{xxing@sjtu.edu.cn}
\address{$^1$Wilczek Quantum Center, School of Physics and Astronomy, Shanghai Jiao Tong University, Shanghai 200240, China \\
$^3$T.D. Lee Institute, Shanghai Jiao Tong University, Shanghai 200240, China\\
$^4$Shanghai Research Center for Quantum Sciences, Shanghai 201315, China}
\date{\today} 
	
	
\begin{abstract}
Using the recently developed covariant Ito-Langevin dynamics, we develop a non-equilibrium thermodynamic theory for small systems coupled to multiplicative noises.  The theory is based on Ito-calculus, and is fully covariant under time-independent nonlinear transformation of variables.   Assuming instantaneous detailed balance, we derive expressions for various thermodynamic functions, including work, heat, entropy production, and free energy, both at ensemble level and at trajectory level, and prove the second law of thermodynamics for arbitrary non-equilibrium processes.  We relate time-reversal asymmetry of path probability to entropy production, and derive its consequences such as fluctuation theorem and non-equilibrium work relation.  For Langevin systems with additive noises, our theory is equivalent to the common theories of stochastic energetics and stochastic thermodynamics.  We also discuss examples of multiplicative noises where the common theories  are inapplicable, but our theory yields correct results.  

\end{abstract}

\maketitle 

\section{Introduction}
\vspace{-3mm}
One of the main themes of modern non-equilibrium physics is to study irreversible processes in small systems, where fluctuations play a dominant role and stochastic descriptions become necessary.   Huge progresses have been achieved in this course in the past few decades~\cite{Seifert-review,Evans-Searles,Jarzynski-review}.  Thermodynamic quantities such as heat, work, and entropy production, were defined at the level of dynamic trajectory~\cite{Sekimoto-book,Seifert-2005}; the deep connection between entropy production and time-reversal asymmetry was discovered~\cite{Maes-2003,Seifert-2005}; a hierarchy of exact identities  known as {\em fluctuation theorems}~\cite{Seifert-review,Searles-1999,Crooks-1999}  and {\em work relations}~\cite{Jarzynski-review}, have been established.  Terms such as {\em stochastic energetics}~\cite{Sekimoto-book} and {\em stochastic thermodynamics}~\cite{Seifert-review} have been coined to describe this very diversified and important field which combines stochastic methods with mechanical and thermodynamic concepts.   More recently serious efforts have been spent to generalize these theories to strongly coupled systems~\cite{Jarzynski-2004-HMF,Jarzynski-2017-strong-coupling,Seifert-2016-strong-coupling,Hanggi-2020-Colloquium,DTX-2021-strong}, system driven by multiplicative noises~\cite{Spinney-Ford-2012-1,Spinney-Ford-2012-2}, systems in moving frames~\cite{Speck2008}, \xing{systems with position-dependent  temperature~\cite{Bo2017,Bo2019,Celani2012}}, as well as quantum systems~\cite{Campisi-2011-Colloquium,Esposito-2009-review-quantum-FT}.   Nonetheless, there does not yet exist a single unified theoretical formulation that is capable of addressing all these issues.  \xing{Additionally, even though formulation of Langevin dynamics in curved space was attempted by Graham~\cite{Graham-covariance-Langevin} more than forty years ago, stochastic thermodynamics of Langevin systems with curved state spaces have not been developed.  }

\vspace{-3mm}

The present work is the third of a sequel dedicated to a  general theory of non-equilibrium statistical physics and thermodynamics using the formalism nonlinear Langevin dynamics and Fokker-Planck dynamics.   In the first paper~\cite{covariant-Langevin-2020}, a covariant theory of nonlinear Langevin dynamics and Fokker-Planck dynamics was developed using Ito-calculus, which are applicable to system driven by multiplicative noises.  One distinguishing feature of the covariant Langevin equation is the appearance of derivatives of kinetic coefficients, i.e., the {\it spurious drift}, which are important to guarantee detailed balance.   Another feature is that differential of slow variables transform not as usual vectors, but according to Ito's formula.  In the second paper~\cite{DTX-2021-strong}, we and collaborator discussed the proper definitions of thermodynamic variables for small Hamiltonian systems that are strongly coupled to environment, such that the interaction is not negligible comparing with the system Hamiltonian.  It was demonstrated that, if the dynamics of bath variables is much faster than that of the system variables, the weak coupling theory of stochastic thermodynamics is also applicable in the strong coupling regime, as long as we identify the fluctuating internal energy as the Hamiltonian of mean force.  This theory should be contrasted with other strong coupling theories, where various thermodynamic quantities need to be defined differently than those in the weak coupling regime.  It was also demonstrated that,  at both trajectory and ensemble levels, heat is negative the change of environmental entropy, conditioned on the system variables.  


In the present paper, we shall combine the results of the previous two works~\cite{covariant-Langevin-2020,DTX-2021-strong} to formulate 
a covariant non-equilibrium thermodynamic theory for small classical systems that are coupled to generally multiplicative noises. It is further assumed that the space of slow variable may be curved, or curvilinear coordinates are used, so that the metic tensor plays a serious role.   As in the previous two works~\cite{covariant-Langevin-2020,DTX-2021-strong}, we shall assume that our system is in contact with a single bath with fixed temperature, and is driven by certain time-dependent control parameter. The system dynamics is described by Ito-Langevin dynamics with white Gaussian noises.  If the the control parameter is fixed, the system converges to a thermal equilibrium which exhibits detailed balance.  (This excludes systems driven by non-conservative forces, and hence do not have a globally defined Hamiltonian.)  Within this setting, we define non-equilibrium work and heat in terms of system variables, both at trajectory level and at ensemble level, and show that the total entropy always increases monotonically, regardless of the initial state of the system.   We also relate time-reversal asymmetry of path probability to entropy production, and prove Crook's Fluctuation Theorem and Jarzynski Equality.   

The main message of this work is that if the metic tensor or the kinetic coefficients depend on system variables, spurious drift shows up in the Langevin equation, and the commonly accepted theories of stochastic energetics and stochastic thermodynamics (which hereafter shall be referred to as {\em the common theories}) need to be replaced by our covariant theory.  This is due to two important reasons. Firstly, in the common theories, Langevin equations are formulated on the idea of force balance.  These equations are generally incorrect when spurious drift play a role.  Secondly, in the common theories, heat is defined, \`a la Sekimoto~\cite{Sekimoto-book}, as the work by friction and random forces. This definition is generally inapplicable to nonlinear Langevin dynamics where friction and reactive force can not be clearly identified.  More seriously, in the presence of spurious drift, Sekimoto's definition of heat leads to violation of the second law of thermodynamics.  By strong contrast, our theory is free of these pathologies.  In Sec.~\ref{sec:example}, we discuss three systems where the common theories fail, whereas our theory gives correct results.  In these examples, the spurious drifts are caused by, respectively, curvilinear coordinates, curved space, and multiplicative noises.  

There are many more sophisticated models of stochastic dynamics where spurious drift shows up, such as rotational diffusion of anisotropic objects and spin dynamics as described by stochastic Landau-Lifshitz-Gilbert equation~\cite{Aron-2014-LLG}.  More interestingly, quantum Markov processes can be described as Brownian motion in Hilbert space~\cite{SSE-book}, which is also curved due to normalization.  Strong interactions between a system and its environment may also lead to state-dependent noise correlations, which also leads to spurious drift.  Hence it is fair to say that in nonequilibrium statistical physics, spurious drifts are rule of thumb rather than exceptions.  We will explore the above mentioned problems in the future.   


The remaining of this work is organized as follows.  In Sec.~\ref{sec:recap}, we adapt the covariant Langevin dynamics developed  in Ref.~\cite{covariant-Langevin-2020} to systems whose slow variables form a Riemannian manifold.  In Sec.~\ref{sec:covariant-Sto-Therm} we define thermodynamic quantities and establish the first and second laws at the ensemble level.  In Sec.~\ref{sec:heat-work-trajectory}, we define thermodynamic quantities at the level of dynamic trajectory, establish the connection between entropy production and time-reversal asymmetry of path probability, derive Crooks Fluctuation Theorem and Jarzynski equality.  In Sec.~\ref{sec:comparison} we discuss the connections between our theory and the classical irreversible thermodynamics as well as Hamiltonian dynamics. We also show that our theory is consistent with the common theories in the case of classical point particles coupled to additive noises.   In Sec.~\ref{sec:example} we discuss three concrete examples of Langevin dynamics with spurious drift, where the common theories fail but our covariant theory give the correct results.  Finally in Sec.~\ref{sec:conclusion} we draw conclusive remarks and outline future directions.   In Sec.~\ref{sec:app} we provide a detailed derivation of the short time transition probability for  nonlinear Langevin dynamics with multiplicative noises.  

\section{Covariant Langevin Dynamics on Riemannian Manifold}
\label{sec:recap}

In this section, we shall reformulate the covariant Langevin theory developed in Ref.~\cite{covariant-Langevin-2020} for systems whose  slow variables form a Riemannian manifold.    

\subsection{Brief review of covariant Langevin dynamics}
 The covariant form of nonlinear Langevin equation with multiplicative Gaussian white noises was derived in Ref.~\cite{covariant-Langevin-2020}:
\ba
&& d x^i(t) + \left( L^{ij} \partial_j U 
- \partial_j L^{ij} \right) dt
=  b^{i \alpha} d W_{\alpha}(t) ,  
\quad\quad\quad
\label{Langevin-def}
\ea
where $\xv = (x^1, \cdots, x^n)$ are the slow variables, $\partial_j = \partial /\partial x^i$, and $d W_{\alpha}(t)$ are the Wiener noises which satisfy:
\ba
&& dW_{\alpha}(t) dW_{\beta}(t) =  \delta_{\alpha\beta} dt.
  \label{Ito-rule}
\ea
In the LHS of Eq.~(\ref{Langevin-def}), $L^{ij}(\xv)$ are {\em the kinetic coefficients}, and $b^{i \alpha}(\xv)$ {\em the noise amplitudes}, both generally depending on the slow variables $\xv$.  The term $- \partial_j L^{ij}  dt$ is called {\em spurious drift}, and shows up when kinetic coefficients depend on $\xv$, i.e., the noises are multiplicative.  Throughout the work, repeated indices are summed over, unless otherwise specified.  In the RHS of Eq.~(\ref{Langevin-def}),  the product  $ b^{i \alpha}(\xv) d W_{\alpha}(t)$ is interpreted in Ito's sense.  For definitions of Ito product and Stratonovich product of stochastic variables, we refer to Ref.~\cite{Gardiner-book}.  The matrix $L^{ij}(\xv) $ is decomposed into a symmetric part $B^{ij}(\xv)$ and an antisymmetric part $Q^{ij}(\xv)$:
\begin{subequations}
\ba
L^{ij}(\xv) &=& B^{ij}(\xv)  + Q^{ij}(\xv), \\
 B^{ij}(\xv) &=&  B^{ji}(\xv) , \\ 
 Q^{ij}(\xv) &=&  -Q^{ji}(\xv).
\ea
 The matrix $b^{i \alpha}(\xv) $ is related to $ B^{ij}(\xv) $ via
\be
b^{i \alpha}(\xv)  b^{j \alpha} (\xv) = 2\, B^{ij}(\xv)
  = L^{ij}(\xv) + L^{ji}(\xv).  
\label{B-b-relation}
\ee
\end{subequations}
All eigenvalues of $ B^{ij}(\xv) $ are non-negative, otherwise either normalization or positivity of probability is violated. 
 
In Ref.~\cite{covariant-Langevin-2020} we define $p(\xv, t) d^n \xv$ to be the probability that the slow variables take value in the infinitesimal volume $d^n \xv$ centered at $\xv$.  Hence $p(\xv, t)$ is the probability density function (pdf) of slow variables.   The function $U(\xv)$ in Eq.~(\ref{Langevin-def}) is called {\em the generalized potential}, which is related to the steady state pdf $p^{\rm SS}(\xv) $ via 
\ba
p^{\rm SS}(\xv) = e^{-U(\xv)}.  
\label{p_SS-U}
\ea 
If the system has a well defined Hamiltonian and is in contact with a single heat bath (as we will assume in this work), the steady state is the thermal equilibrium state.  If the slow variable space is infinite, $U(\xv)$ must be bounded from below and diverge sufficiently fast as $\xv \rightarrow \infty$, in order for $p^{\rm SS}(\xv)$ to be normalizable.  

The covariant form of Fokker-Planck equation associated with the Langevin dynamics (\ref{Langevin-def}) is given by
\ba
\partial_t\,  p(x,t) &=& L_{\rm FP} \, p(x, t) 
\nonumber\\
&=& \partial_i L^{ij} ( \partial_j + (\partial_j U ) ) \, p(x, t). 
\label{FPE-1}
\ea

As demonstrated in Ref.~\cite{covariant-Langevin-2020}, under  nonlinear transformation of variables (NTV) $\xv =(x^i) \rightarrow \xv' = (x'^a )$, $L^{ij}(\xv) $ and $b^{i \alpha}(\xv) d W_{\alpha}$ transform respectively as contra-variant tensor and vector,  whereas $e^{-U(\xv)}d^nx$ and $p(\xv, t)d^nx$ transform as scalars.  In the proof of covariance,  Eq.~(\ref{Ito-rule}) plays an essential role.   Qualitatively speaking, Wiener noise $dW_{\alpha}(t)$ scales as $\sqrt{dt}$, and hence according to Eq.~(\ref{Langevin-def}), $d\xv$ receives contributions from noises of order $\sqrt{dt}$ and from the systematic forces (the second term on LHS) of order $dt$.  As a consequence, to calculate the variation of a function $f(\xv)$ up to order $dt$, we must expand $f(\xv+ d \xv)$ up to order $d\xv^2$:
\ba
d f(\xv) &=& f(\xv + d \xv) - f (\xv)
\nonumber\\
&=& \partial_j f(\xv) \, dx^j 
+ \frac{1}{2}\partial_i \partial_j f(\xv) \, dx^i dx^j. 
\ea  
The quadratic terms $dx^i dx^j$ can be further simplified using Langevin equation (\ref{Langevin-def}) and Eq.~(\ref{Ito-rule}):
\be
dx^i dx^j = \sum_\alpha b^{i \alpha} b^{j \alpha} dt + O(dt^{3/2}),
\label{dxdx-dt}
\ee
where the neglected terms vanish in the continuum limit.  This leads to {\em Ito's formula}~\cite{Gardiner-book}:
\ba  
d f(\xv) =  \partial_j f(\xv) \, dx^j  
 + B^{ij}  \partial_i \partial_j f(\xv)   dt.   
\label{Ito-formula}
\ea
It is important to note that Eq.~(\ref{Ito-formula}) should be understood as an equality of stochastic variables which holds in probability, not just in average. 

\subsection{Langevin dynamics on Riemannian manifold}

The covariant Langevin equation (\ref{Langevin-def}) does not involve metric tensor, hence the range of applicability is not limited to Riemannian manifold.  On the other hand, the pdf $p(\xv,t)$ and $e^{-U(\xv)}$ transform as densities, but not as scalars.  For many physical systems, there is a natural volume measure $dv(\xv)$ defined in the manifold of slow variables, such that the volume of a region $\Omega$ is given by
\be
{\rm Vol}[\Omega] = \int_\Omega   dv(\xv). 
\ee  
The volume measure must be invariant under NTV, which means that in the new coordinates we have $dv'(\xv') \equiv dv(\xv)$. 

For example, if the manifold of slow variables is a Riemann manifold with a metric tensor $ g_{ij}$, it has a natural volume measure 
\ba
dv(\xv) = \sqrt{g(\xv)} \, d^n \xv, 
\label{dv-def}
\ea 
where $g = \det (g_{ij})$ is the determinant of the covariant metric tensor.   
As another example, we consider a classical Hamiltonian system coupled to a heat bath, the set of slow variables consists of canonical coordinates and  momenta $\xv = (\qv, \pv)$.  It is well known that Poisson brackets and Liouville volume measure are preserved by all canonical variable transformations in the phase space.  Hence  the  volume measure, i.e., {\em the Liouville measure}, $d^n \xv = \prod_{i} dp_i dq^i$, is invariant under all canonical variable transformations.  

In the remaining of this work, we shall always assume that the slow variable manifold is Riemannian with metric tensor $g_{ij}(\xv)$, and with an invariant volume measure $\sqrt{g{(\xv)}} \, d^n \xv$.  To apply the theory to Hamiltonian systems, we only need to let $\sqrt{g(\xv)} = 1$, and restrict the nonlinear transformations of variables to canonical transformations.  Throughout the work we shall use the following simplified notation for integration over the invariant measure:
\be
\int_{\xv} \equiv 
 \int \sqrt{g(\xv)} d^n \xv
 =  \int \sqrt{g'(\xv')} d^n \xv'. 
\label{integration-def}
\ee

It is then convenient to define an invariant pdf $\pdf(\xv)$ and  invariant generalized potential $\U(\xv)$ via 
\ba
 \pdf(\xv)  \sqrt{g{(\xv)}} \, d^n \xv
&=& p(\xv)  d^n \xv, \\
 e^{-\U(\xv)}  \sqrt{g{(\xv)}} \, d^n \xv
&=&  e^{-U(\xv)} d^n \xv.  
 \ea
Hence we have the following relations:
\ba
p(\xv) &=& \sqrt{g(\xv)} \, \pdf(\xv), \\
U(\xv) &=& \U(\xv) - \log \sqrt{g(\xv)}.
\ea
Using these, we can rewrite the Langevin equation (\ref{Langevin-def}) and Fokker-Planck equation (\ref{FPE-1}) as 
\ba
&& d x^i + \left( L^{ij} \partial_j  { \U}
- \frac{1}{\sqrt{g}} \partial_j \sqrt{g} L^{ij} \right) dt 
= b^{i \alpha} dW_{\alpha},
\quad\quad\quad \label{Langevin-f}
\\
&& \partial_t \pdf = \frac{1}{  \sqrt{g} } \partial_i   \sqrt{g} L^{ij} 
(\partial_j + (\partial_j { \U})) \pdf
= \LFP \pdf. 
\label{FPE-f}
\ea
We shall call the term $- ({1}/{\sqrt{g}}) (\partial_j \sqrt{g} L^{ij}) dt $ in Eq.~(\ref{Langevin-f}) the {\em spurious drift}, which shows up whenever the kinetic coefficients $L^{ij}$ or the metric determinant $g$ depends on the slow variables $\xv$. 

It is then easy to see that the steady state is
\be
\pdf^{\rm SS} (\xv) = e^{- \U(\xv)}
\label{pdf-SS}
\ee
and satisfies Eq.~(\ref{FPE-f}).    The invariant Fokker-Planck operator $ \LFP$ in Eq.~(\ref{FPE-f}) is 
\ba
\LFP =  \frac {1}{\sqrt{g}} \partial_i  \sqrt{g} L^{ij} 
(\partial_j + (\partial_j { \U})), 
\label{FPO-def-1}
\ea
which is related to $L_{\rm FP}$ defined in Eq.~(\ref{FPE-1}) via
\be
{\mathcal L}_{\rm FP} =  \frac {1}{\sqrt{g}} 
  L_{\rm FP} \sqrt{g}. 
\ee


Now the transformation laws of various components of Eqs.~(\ref{Langevin-f}) and (\ref{FPE-f}) can be obtained from the corresponding results in Ref.~\cite{covariant-Langevin-2020}.  Specifically $dx^i$ transforms according to the Ito's formula:
\ba  
d x' \rightarrow dx'^a =  \frac{\partial x'^a}{\partial x^i} dx^i  
 +\frac{\partial^2 x'^a}{\partial x^i \partial x^j}   B^{ij} dt,   
\label{Ito-formula-dx}
\ea
whilst $\pdf(\xv)$ and $\U(\xv)$ transform as scalars, whereas $b^{i \alpha}(\xv) $ and $L^{ij}(\xv)$ transform respectively as contra-variant vector and tensor: 
 {\begin{subequations}
\label{transform-U-L-b-new}
\ba
\pdf (\xv) &\rightarrow& \pdf'(\xv') = \pdf(\xv). 
\label{covariant-rules-1-new}\\
\U (x )  &\rightarrow& \U'(\xv') = \U (\xv) , 
\label{covariant-rules-U-new} \\
b^{i \alpha}(\xv) &\rightarrow& b'^{a \alpha}(\xv') = \frac{\partial x'_a}{\partial x^i}
b^{i \alpha}(\xv),
\label{covariant-rules-b-new}\\
L^{ij}(\xv) &\rightarrow& (L')^{ab}(\xv')  =
 \frac{\partial x'^a}{\partial x^i} L^{ij} (\xv)
\frac{\partial x'^b}{\partial x^j}. \quad \quad
\label{covariant-rules-3-new}
\ea 
\end{subequations}
The metric transforms as rank two covariant tensor:
\ba
g_{ij} \rightarrow g'_{ab} = \frac{\partial x^i}{\partial x'^a} g_{ij}
\frac{\partial x^j}{\partial x'^b},
\ea
such that $g_{ij}dx^i dx^j = g_{ab} dx'^a dx'^b$ transforms as a scalar.  Using Eq.~(2.20a) of Ref.~\cite{covariant-Langevin-2020}, we can also see that the invariant Fokker-Planck operator, defined in Eq.~(\ref{FPO-def-1})  indeed transforms as a scalar:}
\be
\mathcal L_{\rm FP}'(\xv') = \mathcal L_{\rm FP}(\xv).   
\ee
Note that our notations are consistent with usual tensor analysis, where upper indices and lower indices are used respectively for contra-variant and covariant objects. 

Note that if we choose $L^{ij} = (g^{-1})^{ij}$ to be the contra-variant metric tensor and $ \U (\xv)$ constant, the Fokker-Planck operator (\ref{FPO-def-1})  becomes the covariant Laplacian in Riemannian manifold: $\Delta = { \sqrt{g}^{-1} } \partial_i   \sqrt{g} (g^{-1})^{ij} \partial_j$,  which is  known to be invariant under reparameterization.  

An important but surprising fact about Eq.~(\ref{Langevin-f}) is that neither $dx^i$ nor $- L^{ij} \partial_j ({ \U} - \log  \sqrt{g} ) dt  + \partial_j L^{ij} dt$ transforms as a vector.  But the linear combination $dx^i + L^{ij} \partial_j  ({ \U} - \log  \sqrt{g} )  dt- \partial_j L^{ij} dt$ appearing in the LHS of Eq.~(\ref{Langevin-f}) does behave as a contra-variant vector, and so is the Ito product $b^{i \alpha} d W_{\alpha}$ in the RHS.

In Ref.~\cite{covariant-Langevin-2020}, $b^{i \alpha}$, $L^{ij}$ and $U$ are assumed to be independent of time.  However it is easy to realize that the formalism still works if these functions are time-dependent.  In this work, we shall assume that $b^{i \alpha}$, $L^{ij}, B^{ij} , Q^{ij}$ and $\U$ may  depend on a set of parameters $\lambda$, which may be externally controlled to vary over time.  This is necessary since we aim to study non-equilibrium processes in small systems.   When there is no danger of confusion, we will hide their dependence on $\lambda(t)$.  We shall always assume that the metric tensor ${\mathbf g}$ is time-independent.  There are problems where the metric structure of the slow variables changes with time, such as the diffusion problem on a deformable membrane.  Note also that we only consider NTV independent of time.   In Ref.~\cite{Speck2008}, Speck et. al. use time-dependent variable transformation to study stochastic thermodynamics in moving frames.  We shall not touch this issue here. 


\subsection{Reversibility and detailed balance}
\label{sec:reversible}
As specified in the Introduction, the system being studied has a well defined Hamiltonian and is in contact with a single heat bath.  For fixed control parameter, the system converges to a unique thermal equilibrium state.  The system dynamics then has time-reversal symmetry, whose implications are discussed here.  


As in Ref.~\cite{covariant-Langevin-2020}, we assume that each slow variable has definite time-reversal symmetry.  Hence under time-reversal the slow variables transform as $\xv \rightarrow \xv^*$,  { where $\xv^*= ( \varepsilon_1 x^1, \varepsilon_2 x^2, ..., \varepsilon_n x^n )$ with $\varepsilon_i = + 1, -1$ for even and odd variables respectively.} Take Hamiltonian systems as an example, we have $\xv = (\qv, \pv)$ and $\xv^* = (\qv, -\pv)$, where $\qv$ are the canonical coordinates and $\pv$ the canonical momenta.   For all problems we know the metric tensor is itself invariant under time-reversal:
\begin{subequations}
\label{g-reversal-invariance-1}
\be
\varepsilon_i \,  g_{ij} (\xv^*) \, \varepsilon_j  = g_{ij} (\xv),
\label{g-reversal-invariance}
\ee
where no summation is implied about repeated indices.  Hence the invariant volume measure $dv(\xv)$ is also invariant under time-reversal: 
\ba
{g (\xv^*)}   = {g (\xv)}, \quad dv(\xv^*) = dv(\xv).
\label{g-v-reversal-invariance}
\ea
\end{subequations}

The external control parameter $\lambda$ transforms under time-reversal as $\lambda \rightarrow \lambda^*$.  For a moment we shall assume that $\lambda$ does not vary over time.   Two typical examples where $\lambda$ changes sign under time-reversal are magnetic field and angular velocity.   Note that if $\lambda \neq \lambda^*$, the process with external parameter $\lambda$ is different from that associated with $\lambda^*$.  It is customary to call the process with parameter $\lambda$ the {\em forward process} and that with parameter $\lambda^*$ the {\em backward process}.  The backward process is also described by Langevin equation (\ref{Langevin-f}) and Fokker Planck equation (\ref{FPE-f}), but with $\U(\xv, \lambda), b^{i \alpha}(\xv, \lambda), L^{ij}(\xv, \lambda)$ replaced by  $\U(\xv, \lambda^*), b^{i \alpha}(\xv, \lambda^*), L^{ij}(\xv, \lambda^*)$.    Accordingly, the steady state distributions of the forward and backward processes are 
\begin{subequations}
\label{DB-steady}
\ba
 \pdf_{F}(\xv_1) &=& e^{- \U(\xv_1, \lambda)} , \\
  \pdf_{B}(\xv^*_1) &=& e^{- \U(\xv^*_1, \lambda^*)}.
\ea  
\end{subequations}
Since $(\lambda^*)^* = \lambda$, the backward of the backward process is the forward process itself.  

 Let $\pdf_{F}(\xv_2, t_2; \xv_1, t_1) $ and $\pdf_{ B}(\xv_2, t_2; \xv_1, t_1)$ be the two-time joint pdfs of the forward and backward processes respectively, both assumed in the stationary regime.  Here and below the subscripts in boldface variables $\xv_1, \xv_2 \cdots$ denote time, not the components of slow variables.   These pdfs must have time-translational symmetry, i.e., they can only depend on the time difference $t_2 - t_1$.  A stochastic dynamics is said to to be {\em reversible} if the following relation is satisfied:
\begin{subequations}
\label{DB-condition-1}
\ba
&& \pdf_{F}(\xv_2, t_2; \xv_1, t_1) dv(\xv_2)dv(\xv_1)
\\
&=& \pdf_{ B}(\xv_1^*, - t_1; \xv_2^*, - t_2)
dv(\xv_2^*)dv(\xv_1^*). 
\nonumber
\ea
Because of Eq.~(\ref{g-v-reversal-invariance}),  the volume elements in two sides can be dropped, which leads to
\be
 \pdf_{F}(\xv_2, t_2; \xv_1, t_1)
= \pdf_{ B}(\xv_1^*, - t_1; \xv_2^*, -t_2). 
\label{DB-2-1}
\ee
Because of stationarity, these joint pdfs only depend on $t_2 - t_1$, and Eq.~(\ref{DB-2-1}) is equivalent to 
\ba
 \pdf_{F}(\xv_2, t_2; \xv_1, t_1)
= \pdf_{ B}(\xv_1^*, t_2; \xv_2^*, t_1). 
\label{DB-2-2}
\ea
\end{subequations}

Note also that this condition is symmetric with respect to the exchange of the forward and backward processes, hence the forward process is reversible if and only if the backward is so.  Furthermore, a process may be reversible even if it is not the same as the backward process, i.e. if $\lambda \neq \lambda^*$.   Stationarity together with Eqs.~(\ref{DB-condition-1}) imply that the system is in thermal equilibrium, i.e., the steady states are actually thermal equilibrium states.  

Our definition of reversibility may appear confusing because in quantum mechanics, it is often said that magnetic field (and rotation of frame) breaks time-reversal symmetry.  In this work, however, we shall use the term ``irreversible'' to describe dissipative processes that produce entropy.  Magnetic field and rotation of frame only shift the equilibrium state, but do not lead to a dissipative non-equilibrium state.  Hence, according to our terminology, they do not change the reversible nature of the dynamic process.  

Integrating Eq.~(\ref{DB-2-1}) over $\xv_2$, we obtain $\pdf_{F}(\xv_1,t_1) =   \pdf_{B}(\xv^*_1, -t_2)$.  But these one-time pdfs are given by Eq.~(\ref{DB-steady}).  Hence reversibility implies 
\ba
 \U(\xv^*, \lambda^*) = \U(\xv, \lambda). 
 \label{U-U*-1}
 \ea
Because the processes are Markovian, the two-time pdfs can be written as products of transition probabilities, i.e., conditional pdfs, and one-time pdf's:
\ba
 \pdf_{F}(\xv_2, t_2; \xv_1, t_1) &=&
 \pdf_{F}(\xv_2, t_2| \xv_1, t_1) 
 \pdf_{F}(\xv_1) ,
\\
 \pdf_{ B}(\xv_1^*, - t_1; \xv_2^*, - t_2 ) &=& 
\pdf_{ B}(\xv_1^*, - t_1|  \xv_2^*, - t_2)
\pdf_{ B}( \xv_2^*).  \nonumber \quad
 \ea
Combining these  with Eqs.~(\ref{DB-steady}), we obtain another implication of reversibility:
\ba
\frac{ \pdf_{F}(\xv_2, t_2| \xv_1, t_1) }
{\pdf_{ B}(\xv_1^*, - t_1 |  \xv_2^*, - t_2 ) }
=  e^{- \U(\xv_2, \lambda) + \U(\xv_1, \lambda)}.  
\label{DB-condition-3}
\\\nonumber
\ea
It is important to note that RHS is independent of time, whereas in LHS, the conditional pdfs only depend on time difference $\Delta t = t_2 - t_1$, as well as on the control parameter $\lambda$.   

Generalizing the proof presented in Ref.~\cite{covariant-Langevin-2020}, we can show that Eqs.~(\ref{U-U*-1}) and (\ref{DB-condition-3}) are equivalent to
\begin{subequations}
\label{DB-condition-2}
\ba
 \U(\xv^*, \lambda^*) &=& \U(\xv, \lambda), 
 \label{DB-condition-2-U} \\
 \varepsilon_i L^{ij}(\xv^*, \lambda^*) \varepsilon_j 
 &=& L^{ji}(\xv, \lambda), 
\label{DB-condition-2-L} \\
\int_{\xv} e^{-\U(\xv, \lambda) }&=& 1.  
\label{DB-condition-2-Int} 
\ea
Note that Eq.~(\ref{DB-condition-2-L}) is equivalent to
\ba
\varepsilon_i B^{ij}(\xv^*, \lambda^*) \varepsilon_j 
&=& B^{ij}(\xv, \lambda ),  
\label{DB-condition-2-B} \\
\varepsilon_i Q^{ij}(\xv^*, \lambda^*) \varepsilon_j 
&=& - Q^{ij}(\xv, \lambda). 
 \label{DB-condition-2-Q}
\ea
\end{subequations}
Equations (\ref{DB-condition-2}) are called {\em detailed balance conditions}.  For Markov processes, Eqs.~(\ref{DB-condition-1}), (\ref{DB-N-points}), and (\ref{DB-condition-2}) are all equivalent.  With detailed balance satisfied, the steady state $e^{-U(\xv, \lambda)}$ then describes thermal equilibrium. An alternative proof of these conditions using Green's function will be presented in a separate work ~\cite{DX-coarse-graining}.   Note also that Eq.~(\ref{DB-condition-2-Int}), even though not shown in most works, is important.  It can be shown that if $e^{-U(\xv, \lambda)}$ is not normalizable, the system  converges to either a dissipative non-equilibrium steady state, or some other state that constantly changing with time.  These will be studied in a separate work.

Because of the Markovian property, we can construct $N$-time joint pdfs in the stationary regime from initial pdfs and transition probabilities, both for the forward process and for the backward process:
\begin{widetext}
\vspace{-5mm}
\begin{subequations}
\label{DB-conditional}
\ba
  \pdf_F(\xv_N, t_N; \cdots; \xv_1, t_1; \xv_0, t_0) & =&
   \pdf_F(\xv_N, t_N|\xv_{N-1}, t_{N-1}) 
\cdots
 \pdf_F(\xv_1, t_1|\xv_0, t_0)  \, e^{- \U(\xv_0, \lambda)}, 
\label{DB-conditional-1}
\\
 \pdf_B(\xv_0^*,t_N;  \cdots; \xv_{N-1}^*, t_{1}; \xv_N^*, t_0) 
 & =& \pdf_B(\xv_0^*, t_N|\xv_{1}^*, t_{N-1}) 
  \cdots 
 \pdf_B (\xv_{N-1}^*, t_1|\xv_N^* , t_0) 
 \, e^{- \U(\xv_N^*, \lambda^*)} . 
 \label{DB-conditional-2}
\ea
\end{subequations}
\end{widetext}
 It is understood that $t_N > t_{N-1}> \cdots > t_1 > t_0$ and  that time propagates from right to left in these expressions.  Taking the ratio of Eqs.~(\ref{DB-conditional-1}) and (\ref{DB-conditional-1}), and using Eqs.~(\ref{DB-condition-3}) and (\ref{U-U*-1}) repeatedly, we find 
\ba
\pdf_F(\xv_N, \cdots, \xv_1, \xv_0) 
&=& \pdf_B(\xv_0^*, \xv_1^*,  \cdots, \xv_N^*). \quad\quad
\label{DB-N-points}
\ea
It is useful to think of $( \xv_N, \cdots, \xv_1, \xv_0)$ as a discretized dynamic trajectory, and $(\xv_0^*, \xv_1^*,  \cdots, \xv_N^*)$ as its time reversal.  Equation (\ref{DB-N-points}) then says that, for stationary reversible Markov process,  the probability of dynamic trajectory is invariant under time reversal of both the path and the dynamic protocol, i.e., reversal of the parameter $\lambda$.  In another word, a trajectory in the forward process is equally probable as the backward trajectory in the backward process.   Combining Eqs.~(\ref{DB-conditional}) and (\ref{DB-N-points}), we may also obtain
\ba
\frac{ \pdf_F(\xv_N|\xv_{N-1})  \cdots \pdf_F(\xv_1|\xv_0) }
{ \pdf_B(\xv_0^*|\xv_{1}^*)   \cdots  \pdf_B (\xv_{N-1}^*|\xv_N^* )}
= e^{\U(\xv_0, \lambda) - \U(\xv_N, \lambda)}. 
\quad\quad
\label{DB-N-points-2}
\ea
We note however that Eqs.~(\ref{DB-condition-2}) and (\ref{DB-N-points-2}) are derived by assuming that the parameter $\lambda$ is fixed.  If $\lambda$ change over time, neither the forward process and the backward process can be stationary, yet the detailed balance conditions, Eqs.~(\ref{DB-condition-2}), are still satisfied.  Eqs.~(\ref{DB-N-points}) and (\ref{DB-N-points-2})  however will be replaced by more complicated relations, see Eqs.~(\ref{EPF-stochastic-1}) and (\ref{EPF-Q-1}). 


\section{Thermodynamics at Ensemble Level}
\label{sec:covariant-Sto-Therm} 

{
We shall now study the non-equilibrium thermodynamics associated with the covariant Langevin dynamics (\ref{Langevin-f}) and Fokker-Planck dynamics (\ref{FPE-f}), assuming that the external parameter $\lambda(t)$ is tuned externally as a function of time, and that detailed balance is satisfied for every fixed $\lambda$.  We start with the setting discussed in Ref.~\cite{DTX-2021-strong}, such that at the microscopic level, the total Hamiltonian of the system and the bath is
\ba
H_{\rm tot} = H_\Xv(\xv; \lambda) + H_B(\yv;\xv),
\label{H-tot-decomp}
\ea 
where $\xv,\yv$ are variables of the system and the bath separately.  Unlike in Ref.~\cite{DTX-2021-strong}, however, here we further assume that the external parameter $\lambda$ is coupled to $ H_\Xv(\xv; \lambda)$ but not to $H_B(\yv;\xv)$.  The decomposition of the total Hamiltonian in Eq.~(\ref{H-tot-decomp}) is such that $H_{\Xv}(\xv; \lambda) $  is the Hamiltonian of mean force (HMF) of the system.  Hence the marginal equilibrium pdf of $\xv$ is
\ba
p_\Xv^{\rm EQ}(\xv) = \frac{1}{Z_{\Xv}} e^{- \beta H_\Xv}
= e^{- \beta H_B + \beta F_{\Xv}}. 
\label{pdf-EQ}
\ea
where $F_\Xv(\lambda)$ is the equilibrium free energy:
\ba
F_\Xv(\lambda) = - T \log \int_{\xv} e^{ - \beta H_{\Xv}(\xv, \lambda)}.  
\ea
 We refer the readers to Ref.~\cite{DTX-2021-strong} for detailed construction of this decomposition.  In the limit of time-scale separation, where the dynamics of $\yv$ (bath variables) is much faster than that of $\xv$ (system variables), we expect that, after taking into account fluctuations of $\yv$, the distribution of $\xv$ evolves according to the Fokker-Planck equation (\ref{FPE-f}).  If we follow the dynamic trajectories of $\xv$, then the dynamics is described by the nonlinear Langevin equation (\ref{Langevin-f}), whereas the fast variables behave as Gaussian white noises.  A mathematical derivation from deterministic unitary microscopic dynamics to  mesoscopic Langevin dynamics or Fokker-Planck dynamics can be achieved using projection operator methods.  

Comparing Eqs.~(\ref{pdf-EQ}) with (\ref{p_SS-U}), we find 
\be
\U(\xv, \lambda) = \beta H_{\Xv}(\xv, \lambda) - \beta F(\lambda).
\label{U-F-B}  
\ee
The condition of detailed balance Eq.~(\ref{DB-condition-2-U}) is then translated into
\be
H_{\Xv}(\xv^*, \lambda^*) = H_{\Xv}(\xv, \lambda), 
\quad
F(\lambda^*)  = F(\lambda). 
\tag{2.30a'}\label{DB-condition-2-F}
\ee
Since both Langevin equation (\ref{Langevin-f}) and Fokker-Planck equation (\ref{FPE-f}) depend  on $\U$  only through $\partial_i \U$, we see that the additive constant $\beta F(\lambda)$ in Eq.~(\ref{U-F-B}) can not be determined from the dynamics of slow variables.  Instead it has to obtained from study of the statistical mechanics of the total system consisting of both the slow variables and fast variables.   

 }
 


Thermodynamic quantities are also defined in the same way as in Ref.~\cite{DTX-2021-strong}.  We identify $H_{\Xv} (\xv, \lambda)$ as the {\em fluctuating internal energy} of the system, and its ensemble average as the {\em internal energy}.   Non-equilibrium entropy and free energy of the system are defined as
\begin{subequations}
\label{first-law-ensemble}
\ba
S[\pdf]  &\equiv& - \int_{\xv} \pdf(\xv)  \log \pdf(\xv),\\
 { F}[\pdf(\xv)] &\equiv& \int_{\xv} \pdf(\xv)
\left( H_{\Xv} (\xv, \lambda) + T \log \pdf(\xv)   \right)
\nonumber\\
&=& \left\langle H_{\Xv}(\xv, \lambda) \right\rangle -  T S[\pdf(\xv)]. 
\label{Gibbs-F-def}
\ea
\end{subequations}

Now consider an infinitesimal process where time evolves by $dt$, the parameter $\lambda$ changes by $d \lambda$, and the pdf of slow variables changes by $d \pdf = \LFP \pdf dt$ according to Eq.~(\ref{FPE-f}).   We define the ensemble averaged differential work and heat as
\begin{subequations}
\label{dQ-dW-def-path}
\ba
 \dbar { W} &\equiv& \int_{\xv} \pdf \,d_{\lambda} H_{\Xv}
= \int_{\xv} \pdf \,( \partial_{\lambda} H_{\Xv} ) d\lambda, 
\label{dW-def} \\
 \dbar { Q} &\equiv& \int_{\xv} H_{\Xv} \, d \pdf 
= \int_{\xv}  H_{\Xv} (\LFP \pdf)  \, dt,
\label{dQ-def}
\ea
\end{subequations}
where $d_{\lambda} H_{\Xv} = ( \partial_{\lambda} H_{\Xv} ) d\lambda$ is the differential of $H_{\Xv}$ due to variation of $\lambda$.  {It is shown in Ref.~\cite{DTX-2021-strong} that $\beta \dbar Q$ as defined in (\ref{dQ-def}) equals to negative the variation of environmental entropy, conditioned on the slow variables $\xv$.}    {Note that even though the kinetic matrices $B^{ij}, Q^{ij}$ may also depend on $\lambda$, only the $\lambda$ dependence of $H_{\Xv}$ contributes to the work $\dbar W$.  }  

{
It is important to note that work and heat are defined in Eqs.~(\ref{dQ-dW-def-path}) in terms of Hamiltonian of mean force $H_\Xv$, but not in terms of the generalized potential $\U$.  Nonetheless, $H_\Xv$ and $T\, \U$ differ from each other only by an additive constant $F(\lambda)$, see Eq.~(\ref{U-F-B}).  Because of the particular form of the operator $\mathcal L_{\rm FP}$ (Eq.~(\ref{FPO-def-1})), one easily see via integration by parts that the heat can be equivalently defined in terms of $\U$ as
\be
 \dbar {Q} \equiv T \int_{\xv} \U \, d \pdf 
= T \int_{\xv}  \U (\LFP \pdf)  \, dt. 
\tag{3.6b'}\label{DB-condition-2-F}
\ee
By contrast, the work cannot be rewritten in terms of $\U$ because the constant $ F(\lambda)$ in Eq.~(\ref{U-F-B}) depends on $\lambda$.  The fact that work cannot be fully determined using the information contained in the Langevin equation should not surprise us.  It is in fact shared by the commonly accepted theories of stochastic energetics and stochastic thermodynamics as well.   To explain this issue more clearly, let us consider adding a time-dependent constant potential $V(t) = V_0\,t/T$ to a Hamiltonian system, where $V_0$ is independent of coordinates and momenta.   It of course means that the external agent does work of the same amount to the system.  Yet introduction of $V_0$ does not affect the dynamic equations of the system, since $\partial_q V(t) = \partial_p V(t) = 0$.  As will be shown in Sec.~\ref{sec:comparison-SE-ST}, our definitions of heat and work reduce to those of Sekimoto for the case of additive noises. 
}

The differential of the internal energy is
\ba
d \langle H_{\Xv}(\xv, \lambda) \rangle 
  &=&   \!\! \int_{\xv} \! \pdf \,d H_{\Xv} 
 +\!  \int_{\xv}\! H_{\Xv}  d \pdf
\quad \label{df_B-ave}
\nonumber\\
 &=& \dbar W +  \dbar Q, 
\label{1st-law-1}
\ea
which has the form of {\em the first law of thermodynamics} at the ensemble level.   We can similarly obtain a differential relation for the non-equilibrium free energy ${ F}[\pdf]$ :
\be
\label{dF-Gibbs}
d { F} = \dbar W +  \dbar Q - T \, d S.  
\ee
In Ref.~\cite{DTX-2021-strong}, it is demonstrated that the differential of the total entropy (the joint Gibbs-Shannon entropy of the system variables and bath variables) is
\ba
d S^{\rm tot} &=&  dS - \beta \dbar Q  
\nonumber\\
&=& - dt\, \int_{\xv} (\log p + U) \LFP \pdf.,
\label{dS^tot-1}
\ea
where in the last equality we have used Eqs.~(\ref{dQ-def}) and (\ref{U-F-B}),  as well as:
\ba
d S[\pdf]  &=& - \int_{\xv}  \log \pdf(\xv) \,  d \pdf(\xv), 
\nonumber\\
&=& - dt \int _{\xv}  \log \pdf(\xv) \LFP \pdf. 
\ea

Now inserting Eqs.~(\ref{FPO-def-1}) and (\ref{integration-def}) into Eq.~(\ref{dS^tot-1}), and integrating by parts~\footnote{here we assume natural boundary condition or reflection boundary condition, so that boundary terms all vanish}, we  further have
\ba
\frac{ d S^{\rm tot}}{dt } \!\! &=&  \!\! \int_{\xv} 
 ((\partial_j + \partial_i \U)  \pdf ) 
\frac{B^{ij}}{\pdf}  
( (\partial_j + \partial_j \U)  \pdf ),  
\quad \label{dS^tot-2}
\ea
which is manifestly positive, since the matrix $B^{ij}$ is so.   Combining this with Eqs.~(\ref{dF-Gibbs}) and (\ref{dS^tot-1}), we obtain:
\begin{subequations}
\ba
d S^{\rm tot} &=&  dS - \beta \dbar Q   \geq 0, 
\label{Clausius-ineq}\\
d S^{\rm tot} &=& \dbar W - d { F}   \geq 0. 
\label{dW-dF}
\ea
\end{subequations}
Whilst Eq.~(\ref{Clausius-ineq}) is called the Clausius inequality, Eq.~(\ref{dW-dF}) is usually known as the {\em principle of minimal work}, which says that the minimal work needed for a process is the change of the system free energy.  In classical thermodynamics, these inequalities are valid only for processes starting from and ending at equilibrium states.  In the present theory, however, they are valid for arbitrary non-equilibrium processes.  






{Note that all thermodynamic variables have been constructed using $H_{\Xv} = T\, \U + F$, $\pdf$, as well as the invariant volume measure, which transform as scalars under NTV.  Consequently, all thermodynamic variables and relations discussed in this section  are invariant under NTV.  It is also important to note that the definitions of thermodynamic variables are contingent on the identification of fluctuation internal energy as the HMF $H_{\Xv}$.  If the system is driven by non-conservative forces, there is no unambiguous definition of internal energy.  Then definitions of heat and work will have to been re-evaluated very carefully. This will be discussed in a future work.

\section{ Thermodynamics at Trajectory Level}
\label{sec:heat-work-trajectory}


\subsection{Work and heat at trajectory level}
\label{sec:WQ-trajectory}

{
Consider a small time step $dt$ along a particular trajectory in the forward process, where the slow variables evolve from $\xv$ at time $t$ to $\xv + d\xv$ at time $t + dt$, while $\lambda$ changes to $ \lambda + d\lambda$. In  Ref.~\cite{DTX-2021-strong},  work at trajectory level was defined as the change of  total energy of the universe, averaged over fast variables. It was then proved that the differential work is also the differential of $H_{\Xv}$ due to the change of $\lambda$.  By contrast, heat at trajectory level was defined as negative the energy change of the environment, again averaged over fast variables.  It was then proved that the heat is also the differential of $H_{\Xv}$ due to the change of $\xv$.  In the setting of nonlinear Langevin dynamics, the environmental variables are not explicitly displayed.  Hence work and heat at trajectory level are defined in terms of system variables as
\begin{subequations}
\label{dW-dQ-path-def-1}
\ba
\dbar { \mathcal W} &\equiv& d_{\lambda} H_{\Xv} 
= (\partial_\lambda H_{\Xv})  \, d\lambda
 \nonumber\\
 &=&  
 H_{\Xv} (\xv , \lambda + d \lambda)
 - H_{\Xv} (\xv, \lambda), 
 \label{dW-path-def-1}
\\
 \dbar { \mathcal Q} &\equiv&  d_{\xv} H_{\Xv} \equiv 
   H_{\Xv} (\xv + d\xv, \lambda ) 
- H_{\Xv} (\xv, \lambda).  \quad
\label{dQ-path-def-1}
\ea
\end{subequations} 
}

 Now the differential of the fluctuating internal energy can be decomposed into work and heat at trajectory level:
\ba
d H_{\Xv} &=& H_{\Xv} (\xv + d\xv, \lambda + d \lambda)
- H_{\Xv} (\xv, \lambda) 
\label{dW-path-def}\\
&\approx&  H_{\Xv} (\xv, \lambda + d \lambda)
 - H_{\Xv} (\xv, \lambda) \quad\quad
\nonumber\\
&+& H_{\Xv} (\xv + d\xv, \lambda ) 
- H_{\Xv} (\xv, \lambda) 
+ O(dt^{3/2}) \nonumber\\
&=& \dbar  \mathcal W + \dbar \mathcal Q
+ O(dt^{3/2})  .
\label{dF_B-forward}
\ea
For typical trajectories, the work is of order $dt$ whereas the heat is of order $d\xv \sim dt^{1/2}$.  Hence neglected terms in the above equation are of order $dt^{3/2}$ and hence do not contribute to the continuous limit.  Hence we arrive at the first law at trajectory level:
\ba
d H_{\Xv} &=& \dbar { \mathcal W}  + \dbar { \mathcal Q}
=  d_{\lambda} H_{\Xv} +  d_{\xv} H_{\Xv}. 
\label{1st-law-path-infinites}
\ea
{Both $d_{\lambda} H_{\Xv} $ and $d_{\xv} H_{\Xv} $ transform as scalars under NTV.  }


The definition of heat, Eq.~(\ref{dQ-path-def-1}), can be expanded in terms of $d\xv$.   In view of Eq.~(\ref{dxdx-dt}), however, we should expand up to $d\xv^2$:
\ba
\dbar { \mathcal Q} &=&   (\partial_i  H_{\Xv}) d x^i 
+ \frac{1}{2} ( \partial_i \partial_j  H_{\Xv} )dx^i dx^j
\nonumber \\
&=&   (\partial_i  H_{\Xv}) d x^i 
+ B^{ij}( \partial_i \partial_j  H_{\Xv} )dt, 
\label{df_B-path-1} 
\ea
where we have also used Eqs.~(\ref{B-b-relation}).

In order for the thermodynamics at trajectory level to be consistent with the thermodynamics at ensemble level, we need to show that the ensemble averages of differential work and heat at trajectory level, defined in Eqs.~(\ref{dW-dQ-path-def-1}), equal to the average work and heat we defined in Eqs.~(\ref{dQ-dW-def-path}).  Here ensemble average means averaging both over the pdf $\pdf(\xv, t)$ and over the noise $d W_{\alpha}(t)$, which acts on the system during $(t, t + dt)$.   Note from Eq.~(\ref{dW-path-def}) that $\dbar { \mathcal W}$ does not depend on noise.  Hence to calculate its ensemble average, we only need to multiply Eq.~(\ref{dW-path-def}) by $\pdf(\xv, t) \sqrt{g(\xv)}$ and integrate over $\xv$.  We find Eq.~(\ref{dW-def}) as expected:
\be
\langle \dbar { \mathcal W} \rangle
= \int_{\xv}  \pdf (x, t ) (\partial_\lambda H_{\Xv} ) d \lambda 
 =  d W.  
\ee

 {To calculate the ensemble average of $ d \mathcal Q$, we  express $dx^i$ in Eq.~(\ref{df_B-path-1}) using the Langevin equation (\ref{Langevin-f}), use Eq.~(\ref{U-F-B}) to expresss $\U$ in terms of $H_{\Xv}$, and further average over noise.   The noise term disappears up on averaging because of Ito-calculus, and we obtain:
\ba
&& \bigg[ B^{ij}\big( \partial_i \partial_j H_{\Xv} 
-  \beta (\partial_i H_{\Xv}) (\partial_j H_{\Xv})  \big)
\nonumber\\
&+& (\partial_i H_{\Xv} ) \partial_j L^{ij} 
+ L^{ij} (\partial_i H_{\Xv})(\partial_j \log \sqrt{g}) \bigg] dt.  
\nonumber
\ea
We further multiply the above equation  by $\pdf(\xv, t) \sqrt{g(\xv)}$, then integrate over $\xv$, and find
\ba
  \langle \dbar { \mathcal Q} \rangle
 &=& dt \int_{\xv}  \pdf 
 \Big[ B^{ij}\big( \partial_i \partial_j H_{\Xv} 
-  \beta (\partial_i H_{\Xv}) (\partial_j H_{\Xv})  \big)
\nonumber\\
&  +&  (\partial_i H_{\Xv} ) \partial_j L^{ij} 
+ L^{ij} (\partial_i H_{\Xv})(\partial_j \log \sqrt{g}) \Big].
\ea  }
Further carrying out a few integrations by parts, we find that  $\dbar { \mathcal Q}$ is indeed Eq.~(\ref{dQ-def}), heat at ensemble level:
\ba
\langle \dbar { \mathcal Q} \rangle 
&=& dt \int_{\xv} H_{\Xv} \frac{1}{  \sqrt{g} } \partial_i   \sqrt{g} L^{ij} 
(\partial_j + (\partial_j { \U})) \pdf 
\nonumber\\
&=& dt \int_{\xv} \mathcal{F}_B {\mathcal L}_{\rm FP } \pdf 
= \dbar Q.  
\ea

\subsection{Forward and backward trajectories, protocols, and processes }  

Following the common terminology in stochastic thermodynamics, we shall use {\em protocol} to denote a generic time-dependent external control parameter $\lambda(t)$, and {\em trajectory} to denote a generic dynamic path $\xv(t)$.   We use the terms {\em backward protocol} for $\tilde \lambda(t) = \lambda^*(-t)$, and  {\em backward trajectory} for $\tilde \xv(t) = \xv^*(-t)$.  To make comparison, we shall also call $\lambda(t)$ and $\xv(t)$ respectively the {\em forward protocol} and  the {\em forward trajectory}.  
We call the Langevin dynamics with the forward protocol $\lambda(t)$ the {\em forward process}, and that with the backward protocol $\tilde \lambda(t)$ the {\em backward process}.   Note that if the forward process is defined in the time interval $(t_I, t_F)$, then the backward process is defined in  $(- t_F, - t_I)$.    {In many previous works, both the forward process and backward process are defined in the time intervals $(0,T)$.  The backward trajectory and protocol are then defined as $\xv^*(T-t)$ and $\lambda^*(T-t)$.  Evidently, the choices of time-interval is only a matter of convenience.  }  Strictly speaking, to define a Langevin process, we need to specify the pdf of slow variables at the initial time.  If  the control parameter $\lambda$ is fixed, and the initial pdf is the equilibrium pdf, then the process is stationary, as we discussed in Sec.~\ref{sec:reversible}.  In this section, we shall not assume the system starting from equilibrium.


Corresponding to every infinitesimal step in the forward process which we discussed in Sec.~\ref{sec:WQ-trajectory}, there is a  step in the backward process, where the system goes from $\xv^* + d \xv^*$ to $\xv^*$, and the parameter from $\lambda^* + d \lambda^* $ to $\lambda^*$.   The resulting variation of $H_{\Xv} $ can be analogously decomposed into work  $\left(  \dbar \mathcal W \right)_{\rm bw}$ and heat $\left(  \dbar \mathcal Q\right)_{\rm bw}$:
\ba
\left( d H_{\Xv} \right)_{\rm bw} 
&=&  H_{\Xv} (\xv^* , \lambda^* )
- H_{\Xv} (\xv^*  + d\xv^* , \lambda^*  + d \lambda^* )
\nonumber\\
&=& \left(  \dbar \mathcal Q\right)_{\rm bw}
-  \left(  \dbar \mathcal W \right)_{\rm bw} 
+ O(dt^{3/2}), 
\\
\left(  \dbar \mathcal W \right)_{\rm bw} &\equiv& 
H_{\Xv} (\xv^*, \lambda^*)
-  H_{\Xv} (\xv^* , \lambda^* + d \lambda^*), 
 \quad \label{dW-path-def-1-bw}
\\
\left(  \dbar \mathcal Q\right)_{\rm bw} &\equiv&  
   H_{\Xv} (\xv^* , \lambda^* ) 
- H_{\Xv} (\xv^*+ d\xv^*, \lambda^*). 
\ea
 Here the subscript bw denotes backward.   {Using the detailed balance condition Eq.~(\ref{DB-condition-2-F}), we see that in the backward step,} $\left( d H_{\Xv} \right)_{\rm bw}, \left(  \dbar \mathcal Q\right)_{\rm bw}, \left(  \dbar \mathcal Q\right)_{\rm bw}$ are the negatives of the corresponding quantities in the forward step:
\begin{subequations}
\ba
\left( d H_{\Xv} \right)_{\rm bw} &=& 
- \left( d H_{\Xv}  \right)_{\rm fw}, \\
\left(  \dbar \mathcal W \right)_{\rm bw} &=&
- \left(  \dbar \mathcal W\right)_{\rm fw}, \\
\left(  \dbar \mathcal Q\right)_{\rm bw} &=&
- \left(  \dbar \mathcal Q\right)_{\rm fw}, 
\ea
\label{F-W-Q-backward-forward}
\end{subequations}
where quantities in RHS with subscript fw refer to the forward process and are given in Eqs.~(\ref{dW-dQ-path-def-1}) and (\ref{dF_B-forward}).  
 
Let us now consider a trajectory of finite duration.  The total heat and work along the forward trajectory $\xv(t)$ in the forward process can be obtained by integrating $\dbar \mathcal Q$ and $ \dbar \mathcal W$ along the path:
\begin{subequations}
\label{Q-W-integrals}
\ba
\mathcal W[\xv(t), \lambda(t) ] &=& 
\int_{t_I}^{t_F}  \dbar \mathcal W,\\
\mathcal Q[\xv(t),  \lambda(t) ] &=&
 \int_{t_I}^{t_F}\dbar \mathcal Q, 
\ea 
\end{subequations}
 The integrated first law takes the form:
\ba
\Delta H_{\Xv} &=&
H_{\Xv}(\xv(t_F), \lambda(t_F)) 
- H_{\Xv}(\xv(t_I), \lambda(t_I)) 
\nonumber\\
&=&  { \mathcal W}  [\xv(t), \lambda(t) ]  
+  { \mathcal Q}[\xv(t), \lambda(t) ]  .  
\label{1st-law-path-finite}
\ea
Note that we have displayed the dependence of $\mathcal W, \mathcal Q$ on trajectory and protocol.  

Let $\mathcal Q[ \tilde \xv(t),\tilde \lambda(t)] $ and $\mathcal W[ \tilde \xv(t),\tilde \lambda(t)]$ be the  {total heat and work} along the backward trajectory $\tilde \xv(t)$ of the backward process, which can be obtained by integrating $\left(  \dbar \mathcal Q\right)_{\rm bw}, \left(  \dbar \mathcal Q\right)_{\rm bw}$ along the backward trajectory.  Using Eqs.~(\ref{F-W-Q-backward-forward}) and (\ref{Q-W-integrals}), we easily find 
\begin{subequations}
\label{F-W-Q-backward-forward-tot}
\ba
\mathcal Q[ \tilde \xv(t),\tilde \lambda(t)] &=& 
- \mathcal Q[\xv(t) ,  \lambda(t) ] , \\
\mathcal W[[ \tilde \xv(t),\tilde \lambda(t)]  &=& 
- \mathcal W[\xv(t),  \lambda(t)  ] .
\ea 
\end{subequations}
The meaning of Eqs.~(\ref{F-W-Q-backward-forward}) is that work and heat are both odd under simultaneous reversal of trajectory and dynamic protocols.

\subsection{Path probability}

We shall now define probability density of dynamic trajectory.  To circumvent the difficulty associated with infinite dimensional space of dynamic trajectories, we discretize the time variable, i.e., use the trick of time-slicing.  We divide the interval $(t_I, t_F )$ into $N \gg 1$ infinitesimal steps with duration $dt = (t_F- t_I)/N$.  In $k$-th step, the system goes from $\xv_{k-1} = \xv(t_{k-1})$ at time $t_{k-1}$ to $\xv_{k} = \xv(t_k)$ at time $t_k$, whereas the parameter changes from $\lambda_{k-1} = \lambda(t_{k-1})$ to $\lambda_k = \lambda (t_k)$.   We can now approximate the forward trajectory $\xv(t)$ as a discrete sequence of states, which we denote as $\gamma$:
\begin{subequations}
\be
\gamma: \quad \begin{pmatrix} \xv_N \\ t_N \end{pmatrix}
\leftarrow 
\cdots
\leftarrow 
\begin{pmatrix} \xv_{k} \\ t_{k} \end{pmatrix}
\leftarrow 
\cdots
\leftarrow 
\begin{pmatrix} \xv_{0} \\ t_{0} \end{pmatrix}, 
\label{path-discrete}
\ee
where $t_N = t_F, t_0 = t_I$, and the arrow indicates propagation of time.   We similarly approximate the backward trajectory $\tilde \xv(t) = \xv^*(-t)$ as the reversed discrete sequence, and denote it as $\tilde \gamma$:
\be
\tilde \gamma: \quad
\begin{pmatrix} \,\, \xv_N^* \\\!   - t_{ N}    \!\! \end{pmatrix}
  \rightarrow 
\cdots
\rightarrow 
\begin{pmatrix} \,\,\xv_{k}^* \\  \!  - t_{k} \!\! \end{pmatrix}
 \rightarrow  
\cdots
 \rightarrow 
\begin{pmatrix} \,\, \xv_{0}^* \\  \! - t_{0} \!\! \end{pmatrix}, 
\label{path-discrete-backward}
\ee
\label{path-discrete-tot}
\end{subequations}
where time propagates from left to right.  Note that the initial state of the backward trajectory $\xv_N^*$ is  time-reversal of the final state of the forward trajectory, whereas the final state of the backward trajectory $\xv_0^*$ is  time-reversal of the initial state of the forward trajectory.  

We further define the invariant volume measure of the forward trajectory as
\ba
D\gamma &\equiv& d v_N \cdots dv_1 dv_0 
\nonumber\\
&=& d v(\xv_N) \cdots dv(\xv_1) dv(\xv_0),  
\ea 
where $dv(\xv)$ is defined in Eq.~(\ref{dv-def}).      We then define the differential probability of the discretized path $\gamma$ in the forward process as the N-time joint pdf:
\ba
\pdf_F[\gamma ] D\gamma &\equiv& 
  \pdf_F(\xv_N, t_N; \cdots; \xv_1, t_1; \xv_0, t_0)
\nonumber\\
&\times& d v_N \cdots dv_1 dv_0. 
\label{path-prob-discrete}
\ea
Similar to Eqs.~(\ref{DB-conditional}), the N-time joint pdf can be decomposed as:
\ba
  \pdf_F(\xv_N, t_N; \cdots; \xv_0, t_0)  
&=& \pdf_F(\xv_N, t_N|\xv_{N-1}, t_{N-1}) 
  \nonumber \\
   &\times & \cdots  \cdots\cdots \cdots\cdots\cdots \nonumber\\
 &\times & \pdf_F(\xv_1, t_1|\xv_0, t_0)  
\label{path-prob-discrete-1}\\
 & \times& \pdf_{F} (\xv_0, t_0).  \quad
\nonumber
\ea
Unlike in Eqs.~(\ref{DB-conditional}), however, here the initial pdf $ \pdf_{F0} (\xv_0,t_0)$ remains arbitrary.   

Let us further introduce the notation  $\gamma_0 = \xv_0$ to denote the initial state of the forward trajectory, and $\pdf_F(\gamma_0) = \pdf_{F} (\xv_0, t_0)$ its pdf.  The product of $N$ transition probabilities in Eq.~(\ref{path-prob-discrete-1}) can be understood as the conditional  pdf of the forward trajectory, given its initial state $\gamma_0 = \xv_0$:
\label{path-prob-1-1}
 \ba
\pdf_F [\gamma | \gamma_0] &=& 
\pdf_F(\xv_N, t_N|\xv_{N-1}, t_{N-1}) 
 \cdots 
  \nonumber\\ 
&\times & \pdf_F(\xv_1, t_1|\xv_0, t_0).
 \label{path-prob-discrete-1-cond}
\ea 
The unconditional path pdf can then be expressed as
\ba
\pdf_F [\gamma] &=& 
\pdf_F [\gamma |  \gamma_0] \,\pdf_F(\gamma_0). 
\label{p-path-cond-uncond-F}
\ea

The same things can be done for  the discretized backward trajectory Eq.~(\ref{path-discrete-backward}) of the backward process.   The invariant volume measure of the backward trajectory is
\ba
D \tilde \gamma &\equiv& 
dv_0^* dv_1^* \cdots dv_N^*
\nonumber\\
&=& dv(\xv_0^*)dv(\xv_1^*)  \cdots  d v(\xv_N^*)
= D\gamma,  
\ea 
where we have used the symmetry Eq.~(\ref{g-v-reversal-invariance}).  The counterpart of Eqs.~(\ref{path-prob-discrete}) is
\ba
\pdf_B[ \tilde\gamma ] D\tilde \gamma &\equiv& 
  \pdf_B(\xv_0^*, - t_0; \xv_1^*, - t_1\cdots; \xv_N^*, - t_N)
\nonumber\\
&\times& dv_0^* dv_1^* \cdots dv_N^*. 
\label{path-prob-discrete-bw}
\ea
The counterpart of Eq.~(\ref{path-prob-discrete-1}) is 
\ba
\pdf_B(\xv_0^*, - t_0; \cdots \xv_N^*, - t_N)  
&=& \pdf_B(\xv_0^*, - t_0|\xv_{1}^*, -t_{1}) 
  \nonumber\\
   &\times & \cdots  \cdots\cdots \cdots\cdots\cdots \nonumber\\
 &\times & \pdf_B (\xv_{N-1}^*, - t_{N-1}|\xv_N^*, - t_N)  
\nonumber\\
 & \times& \pdf_{B} (\xv_N^*, - t_N).  \quad
\label{path-prob-discrete-1-bw}
\ea
 Furthermore introducing the notation $\tilde \gamma_0 = \xv_N^*$ for the initial state of the backward trajectory, and $\pdf_B( \tilde \gamma_0 )$ we have the counterparts for Eqs.~(\ref{path-prob-discrete-1-cond}) and (\ref{p-path-cond-uncond-F}):
\ba
\pdf_B [\tilde \gamma | \tilde \gamma_0] &=& 
\pdf_B(\xv_0^*, -t_0|\xv_{1}^*, -t_1)   \cdots 
\nonumber\\
&\times& \pdf_B (\xv_{N-1}^*, -t_{N-1}|\xv_N^*, -t_N ) .
\label{path-prob-discrete-2-cond} \\
\pdf_B [\tilde \gamma ] &=& 
\pdf_F [\tilde\gamma | \tilde \gamma_0] \,\pdf_B( \tilde \gamma_0 ).
\label{p-path-cond-uncond-B}
\ea

%
%





\subsection{Detailed Fluctuation Theorem} 
\label{sec:Heat Formula}
Recall that in Sec.~\ref{sec:reversible} we have derived Eq.~(\ref{DB-condition-3}) by assuming that $\lambda$ is fixed. If $t_2 - t_1 =dt$ is small, $d\xv = \xv_2 - \xv_1 \sim \sqrt{dt}$ is also small for typically trajectories.  Hence  the exponent in the RHS of Eq.~(\ref{DB-condition-3}) can be rewritten as:
 \be
 - d_{\xv} \U(\xv, \lambda) 
 = - \beta \, d_{\xv} H_{\Xv} (\xv, \lambda)
  = - \beta\dbar \mathcal Q, 
  \label{dQ-dU}
 \ee
 where we have used  Eqs.~(\ref{U-F-B}) and (\ref{dQ-path-def-1}).  Hence Eq.~(\ref{DB-condition-3}) can be rewritten into 
\ba
 \frac{\pdf_{F}(\xv + d\xv, t+ dt | \xv, t ) }
{\pdf_{ B}(\xv^*, -t| \xv^* + d\xv^*, -t-dt)} 
= e^{-\beta d \mathcal Q}. 
\label{DB-condition-3-new}
\ea

Now consider the case that $\lambda(t)$ is varied continuously, both in the forward process and in the backward process.  There comes the issue now when should $\lambda(t)$ be evaluated in Eq.~(\ref{dQ-dU}).  The answer is very simple. In the time step $dt$, $d \lambda \sim dt$, but $\dbar \mathcal Q$ is already of order $dt^{1/2}$.  Hence change of time where $\lambda(t)$ is evaluated only leads to correction of Eq.~(\ref{dQ-dU}) at the order of $dt^{3/2}$, which is negligible in the continuum limit. It does not matter when we evaluate $\lambda(t)$.  
 

In Appendix, we explicitly calculate the short time transition probabilities $\pdf_{F}(\xv + d\xv, t+ dt | \xv, t ) $ using the covariant Langevin equation (\ref{Langevin-f}), and verify Eq.~(\ref{DB-condition-3-new}).  The formula of short-time transition probability obtained in Appendix is useful for construction of path-integral representation of Langevin dynamics in curved space or driven by multiplicative noises .  

Let us  apply Eq.~(\ref{DB-condition-3-new}) to a pair of evolution steps in the forward and backward processes, as shown in Eqs.~(\ref{path-discrete-tot}).   The heat associated with the forward  step is $ \U(\xv_{k+1}, \lambda(\tau_k)) - \U(\xv_k, \lambda(\tau_k)) = d \mathcal Q_k$, where $\tau_k$ is an arbitrary point between $t_k, t_{k+1}$, whose precise value does not matter for the reason explained above.  We then obtain:
\be
 \frac{\pdf_{F}(\xv_{k+1}, t_{k+1} | \xv_k, t_k )}
{\pdf_{ B}(\xv_k^*, -t_k| \xv_{k+1}^*, - t_{k+1})} 
= e^{-\beta d {\mathcal Q}_k }.  
\label{EFP-1step-2}
\ee
We can now calculate the ratio of Eqs.~(\ref{path-prob-discrete-1-cond}) and (\ref{path-prob-discrete-2-cond}). Using Eq.~(\ref{EFP-1step-2}) for every step, we find 
\ba
 \frac{\pdf_F [\gamma |\gamma_0] }
{\pdf_B [\tilde \gamma |\tilde \gamma_0] }
= e^{- \beta \mathcal Q[\gamma, \lambda ] }
= e^{ \beta \mathcal Q[\tilde \gamma, \tilde \lambda ] },
\label{EPF-Q-1} \quad\quad
\ea
where $Q[\gamma, \lambda ],  \mathcal Q[\tilde \gamma, \tilde \lambda ]  $ are respectively the total heat absorbed by the system along the forward/backward trajectory in the forward/backward process, which depends both on the trajectory $\gamma, \tilde \gamma$ and on the protocol $\lambda, \tilde \lambda$.  Equations~(\ref{EPF-Q-1}), or its short time version Eq.~(\ref{DB-condition-3-new}),  play a fundamental role in the theory of stochastic thermodynamics, has often been called the {\em Detailed Fluctuation Theorem}.   Note that this result is a consequence of detailed balance, and is valid for arbitrary trajectory $\gamma$.   {In some previous works, when studying general non-linear Langevin dynamics, Eq.~(\ref{EPF-Q-1}) or (\ref{DB-condition-3-new}) was often treated as definition of heat.  Such a definition however must be carefully dealt with, as its connection to  change of bath entropy is not clear.  In our covariant, however, the equivalence of Eq.~(\ref{EPF-Q-1}) and Eq.~(\ref{dQ-path-def-1}) is explicitly established, and in Ref.~\cite{DTX-2021-strong}, it was demonstrated that $- \beta d_{\xv} H_\Xv$ is indeed the entropy change of environment. 
}

Combining Eqs.~(\ref{EPF-Q-1}), (\ref{p-path-cond-uncond-F}) and (\ref{p-path-cond-uncond-B}), we obtain 
\ba
\log \frac {\pdf_F [\gamma]}{\pdf_B [\tilde \gamma ] }
= - \beta  \mathcal Q[\gamma , \lambda] 
+ \log \pdf_{F}( \gamma_0 )
- \log \pdf_B( \tilde \gamma_0 ) ,
\nonumber\\
\quad \label{EPF-stochastic-1}
\ea
which is valid for arbitrary initial pdfs of the forward and backward processes.  Since $-\mathcal Q[\gamma , \lambda] $ is the energy transfer from the system to the heat bath, $- \beta  \mathcal Q[\gamma, \lambda ] $ is the entropy change of the heat bath, conditioned on the slow variables (assuming, of course, that the bath is always in equilibrium).    Seifert calls $-\log \pdf(\xv)$ {\em the stochastic entropies} of the system at state $\xv$, whose ensemble average gives the usual Gibbs-Shannon entropy.  If one  further choose $ \pdf_{B}(\tilde \gamma_0) = \pdf_{B} (\xv^*, - t_N) = \pdf_{F} (\xv, t_N)$, which means that the initial pdf of the backward process is chosen to be the time-reversal of the final pdf of the forward process, then $ \log \pdf_{F}( \gamma_0 )
- \log \pdf_B( \tilde \gamma_0 )$ is the change of the system stochastic entropy along the trajectory $\gamma$, and the RHS of Eq.~(\ref{EPF-stochastic-1}) may be interpreted as the change of total entropy of the universe along the dynamic trajectory $\gamma$ in the forward process.  
%



\subsection{Crooks and Jarzynski}
\label{sec:FT}
The derivations of Crooks Fluctuation Theorem and Jarzynski equality (Integrated Fluctuation Theorem) from the Detailed Fluctuation Theorem Eq.~(\ref{EPF-stochastic-1}), are standard and have been discussed in many works.  Nonetheless, to make the work self-contained, we briefly present these derivations using our own notations.  Suppose in the forward protocol, the control parameter starts from $\lambda (t_I) = \lambda_I$ and ends at $\lambda (t_F) = \lambda_F$.  In the backward protocol, then the control parameter starts from $\tilde \lambda(-t_F) =  \lambda_F^*$  and ends at $\tilde \lambda(- t_I) = \lambda_I^*$.  

Furthermore, we assume that both in the forward process and in the backward process, the system starts from thermal equilibrium with respect to the control parameter at that moment. Let a generic trajectory $\gamma$ have initial state $\gamma_0 = \xv_I$ and final state $\xv_F$, then $\tilde \gamma_0 = \xv_F^*$, and we have
\begin{subequations}
\label{p_F-B-0}
\ba
\pdf_{F}(\gamma_0) &=& e^{\beta F(\lambda_I) 
- \beta H_{\Xv}(\xv_I, \lambda_I)}, 
\label{p_F-0} \\
 \pdf_{B}(\tilde \gamma_0) &=& e^{\beta F(\lambda_F^*)
 - \beta H_{\Xv}(\xv_F^*, \lambda_F^*)}
 \nonumber\\
 &=& e^{\beta F(\lambda_F)
 - \beta H_{\Xv}(\xv_F, \lambda_F)} ,
 \label{p_B-0} 
\ea
\end{subequations}
where in the last equality we have used detailed balance, Eq.~(\ref{DB-condition-2-F}).    Substituting Eqs.~(\ref{p_F-0}) and (\ref{p_B-0}) back into Eq.~(\ref{EPF-stochastic-1}), and using Eq.~(\ref{1st-law-path-finite}), we obtain 
\ba
\log \frac{\pdf_F[\gamma]}  {\pdf_B[\tilde \gamma ]}
= {\beta {\mathcal W}[\gamma, \lambda ] 
- \beta \Delta F },
\label{EPF-2}
\ea
where $\Delta F \equiv F(\lambda_F) - F(\lambda_I)$ is the difference of Gibbs free energy between the final equilibrium state and the initial equilibrium state.  

One may be attempted to think of Eq.~(\ref{EPF-2})  as  the entropy increase of the universe along the trajectory $\gamma$, from $t_I$ to $t_F$.  But this is not correct in general.  The entropy change of the universe along $\gamma$ can be written as 
\ba
- \beta \mathcal Q[\gamma, \lambda]
 - \log \pdf_F (\xv_F, t_F) + \log \pdf_F(\xv_I, t_I), 
 \label{dS-tot-gamma}
\ea
where the first term is the entropy change of the environment, conditioned on the system variables, and the rest two terms are the change of stochastic entropy of the system.  For a genetic dynamic protocol, the system is off equilibrium at $t_F$, and $\pdf_F (\xv_F, t_F)$ is not Gibbs-Boltzmann.  Hence Eq.~(\ref{dS-tot-gamma}) is not the same as $\beta {\mathcal W}[\gamma, \lambda ]  - \beta \Delta F$.  Nonetheless, if the protocol is such that at $t_F$ the system already equilibrates, then Eq.~(\ref{dS-tot-gamma}) equals to $\beta {\mathcal W}[\gamma, \lambda ]  - \beta \Delta F$, and can be understood as the total entropy change of the universe as the system evolves along the trajectory $\gamma$.  
%
%

Note that the initial conditions of the forward and backward processes, Eqs.~(\ref{p_F-B-0}), are chosen such that they are symmetric with respect to reversal.  In another word, the backward of the backward process is the forward process.   Defining two pdfs as follow:
\begin{subequations}
\ba
p_F(\sigma) &\equiv& \int D \gamma \, \pdf_F(\gamma) \,\,
\delta \left( \sigma - \log \frac{\pdf_F[\gamma]}{\pdf_B[\tilde \gamma]} \right),
\quad \\
p_B(\sigma) &\equiv& \int D\tilde \gamma \, \pdf_B(\tilde\gamma)\,\,
\delta \left( \sigma - \log \frac{\pdf_B[\tilde \gamma]} {\pdf_F[\gamma]} \right).
\ea
\end{subequations}
If the protocols are such that, both in the forward process and in the backward process, the system already equilibrates in the final state, then $p_F(\sigma), p_B(\sigma) $ can be understood as the pdf of entropy production in the forward and backward processes, respectively.

%

Using a  theorem proved by van der Broeck and Cleuven~\cite{Broeck-Cleuren-comment-2007}, one can prove
\be
\frac{p_F(\sigma)}{p_B(-\sigma) } = e^{-\sigma} . 
\label{CFT}
\ee
But according to Eq.~(\ref{EPF-2}), $\log {\pdf_F[\gamma]}/{\pdf_B[\tilde \gamma]}$ is linearly related to the work ${\mathcal W}[\gamma, \lambda ]$ along $\gamma$, hence we see that the probability distribution of work ${\mathcal W}[\gamma] $ obeys the Crooks Fluctuation Theorem:
\be
\frac{ p_F(w)}{ p_B(- w)} = e^{-\beta \Delta { F} + \beta w} .
\ee
 Multiplying both sides by $ p_B(- w) e^{- \beta w} $ and integrating over $w$, we obtain Jarzynski work relation:
\be
\langle e^{ - \beta w} \rangle_F = 
\int dw \, e^{- \beta  w} p_F(w) = e^{-\beta \Delta F}. 
\label{JE}
\ee

 {
 \section{Connections with some previous theories}
 \label{sec:comparison}
In this section, we show that both classical irreversible thermodynamics and  Hamiltonian dynamics can be understood as special limits of our covariant theory.  We also show that for weakly damped classical mechanical systems, our theory is consistent with the common theories of stochastic energetics and stochastic thermodynamics.  

\subsection{Deterministic limit and classical irreversible thermodynamics}
\label{sec:irrev-TM}
As discussed in Ref.~\cite{covariant-Langevin-2020}, there are  two possible scenarios where a deterministic limit of our covariant Langevin dynamics can be obtained: (i) the thermodynamic limit, where $\U$ becomes extensive, whereas both the spurious drift and the noise terms are subextensive; (ii) the low temperature limit, where $\U = \beta H_{\Xv} - \beta F$ becomes large comparing with the spurious drift and noises.  In either case, the spurious drift and noises can be dropped and the covariant Langevin equation (\ref{Langevin-f}) reduces to a deterministic equation: 
 \ba
\dot x^i = -  L^{ij} \partial_j \beta H_\Xv(\xv, \lambda),
\label{irrev-thermodynamics}
\ea 
where $H_\Xv$ is the Hamiltonian of mean force as defined in Eq.~(\ref{U-F-B}).  Up to an additive constant, $- \beta H_\Xv(\xv, \lambda)$ can be understood as the entropy of the universe, including the system and all its environment.  Hence $ - \beta \partial_j H_\Xv = \partial S^{\rm tot}/\partial x^j$ is the {\em affinity} corresponding to the thermodynamic variable $x^i$, whereas $\dot x^i$ is the associated {\em flux}.  Equation (\ref{irrev-thermodynamics}) then becomes the basic equation of irreversible thermodynamics~\cite{Callen-book}, which linearly relates affinities to fluxes, and dictates how thermodynamic variables $\xv$ relax towards equilibrium.  Correspondingly, $ L^{ij}$ are precisely the kinetic coefficients in the classical irreversible thermodynamics, which are usually assumed to be constants.  The detailed balance conditions  (\ref{DB-condition-2}) on the kinetic coefficients then become 
 the famous {\em Onsager-Casimir reciprocal relations}.  

Note however, in classical irreversible thermodynamics, $\dot x^i$ transforms as usual contra-variant vectors, whereas in the covariant Langevin theory, $d x^i$ transforms according to Ito-formula, Eq.~(\ref{Ito-formula-dx}).  The difference between these two transformations is due to the last term in the RHS of Eq.~(\ref{Ito-formula-dx}), which is proportional to $B^{ij}$.   It becomes negligible in the deterministic limit, for the same reason that the spurious drift becomes negligible.  

For the particular case of a Hamiltonian system weakly damped by the environment, we have $\xv = (\qv, \pv) = (q^1, \cdots, q^n, p^1, \cdots, p^n)$, and 
\ba
&& H_\Xv(\qv,\pv) =  K(\pv) + V(\qv) -F , 
\quad\quad\label{underdampedU}
\\
&& \Qm = \begin{pmatrix} 
0 & - \mathbf{I}_n\\  \mathbf{I}_n& 0
\end{pmatrix}, 
\quad
 \Bm = \begin{pmatrix} 
0 & 0 \\ 0 & \boldsymbol \gamma
\end{pmatrix},
\ea
where $K(\pv), V(\qv)$ are respectively the kinetic and potential energies, whilst $\boldsymbol \gamma$ is a $n \times n$ symmetric positive constant matrix.  Equation (\ref{irrev-thermodynamics}) then reduces to 
\ba
\dot q^i &=& \frac{\partial K}{\partial p^i}, \\
\dot p^i &=& -  \frac{\partial V}{\partial q^i}
- \gamma^{ij}  \frac{\partial K}{\partial p^i}.
\ea
These equations are identical to Eqs.~(121.2) and (121.5) of Landau \& Lifshitz~\cite{LL-vol5}.  



\subsection{Unitary dynamics and Hamiltonian dynamics}
As shown in Ref.~\cite{covariant-Langevin-2020}, the unitary limit of our covariant Langevin dynamics is reached by the limit $B^{ij} = 0$, where Eq.~(\ref{Langevin-f}) becomes: 
 \ba
d  x^i + \left(  Q^{ij} \partial_j  { \U}
- \frac{1}{\sqrt{g}} \partial_j \sqrt{g} Q^{ij} \right) dt
= 0.  
\label{Langevin-f-unitary}
 \ea
 In this limit, there is no energy dissipation, and the dynamics becomes deterministic and unitary.  The precise meaning of unitarity is discussed in detail in Ref.~\cite{covariant-Langevin-2020}.    
 
 Recall in Riemann manifold covariant derivatives of tensor fields are defined as:
 \ba
\nabla_k A^{ij} 
= \partial_k A^{ij} + \Gamma^i_{mk} A^{mj} 
+ \Gamma^j_{mk} A^{im},
 \ea
 where $\Gamma^i_{mk}$ are the Christoffel symbols:
 \ba
 \Gamma^i_{kl} = \frac{1}{2} g^{im} 
 \left( 
\partial_l g_{mk} + \partial_k g_{ml} - \partial_m g_{kl} 
 \right),
 \ea
which have the following properties:
\ba
\Gamma^j_{kj} = \partial_k \log \sqrt{g}.  
\ea 
Using these results we can easily show that 
\ba
\nabla_j Q^{ij} &=& \partial_j Q^{ij} 
+ \frac{1}{\sqrt{g}} Q^{ij} \partial_j \sqrt{g}
\nonumber\\
&=&  \frac{1}{\sqrt{g}} \partial_j \sqrt{g}Q^{ij}.
\label{nabla-Q-ij}
\ea
Hence Eq.~(\ref{Langevin-f-unitary}) can be written as
\ba
\dot x^i = v^i(\xv) = -  Q^{ij} \nabla_j  { \U}
+ \nabla_j Q^{ij}  .  
\label{Langevin-f-unitary-1}
\ea
For reason which will become clear below, we assume 
\be
 (\nabla_i Q^{ij} ) (\nabla_j  { \U} ) = 0.  
 \label{assumption-Q-U}
\ee
As a special example, we consider Hamiltonian dynamics with canonical variables.  $Q^{ij}$ is just the constant symplectic matrix, and the Christoffel symbols are zero, hence $\nabla_i Q^{ij}  = 0$, and Eq.~(\ref{assumption-Q-U}) is trivially satisfied.  
 
Recall that $ \nabla_j Q^{ij}$ is a contra-variant vector field, and for any contra-variant vector field we have
 \be
 \nabla_i v^i = \frac{1}{\sqrt{g}} \partial_i \sqrt{g} v^i. 
 \label{nabla-v-i}
 \ee
 Combining this with Eq.~(\ref{nabla-Q-ij}), we find 
 \ba
  \nabla_i \nabla_j Q^{ij} = 0.  
 \ea
Now take the covariant divergence of Eq.~(\ref{Langevin-f-unitary-1}) and use the Leibniz rule of covariant derivatives as well as $\nabla_i \nabla_j \varphi =  \nabla_j \nabla_i \varphi$, we find 
 \be
   \nabla_i \dot x^i  =     \nabla_i  v^i(\xv)
=   -  (\nabla_i Q^{ij} ) (\nabla_j  { \U} )
 =0, 
 \ee
 where the last step follows from our assumption Eq.~(\ref{assumption-Q-U}).  Hence the flow defined by Eq.~(\ref{Langevin-f-unitary-1}) is incompressible, i.e., the space volume of slow variable  is conserved by the unitary dynamics.  The conservation of phase space volume in Hamiltonian dynamics is known as the Liouville theorem. 
 
We can also show that the generalized potential $\U$ is conserved by the dynamics:
\ba
\frac{d \U}{dt} &=& (\partial_i \U) \dot x^i 
\\
&=& -  (\nabla_i \U)Q^{ij} (\nabla_j  { \U})
+ (\nabla_i  { \U} )(\nabla_j Q^{ij} )  
 = 0. \nonumber
\ea
Similarly we can  prove that the Gibbs-Shannon entropy $S[\pdf]$ is also conserved.  Taking time-derivative of the entropy, we find
\ba
\frac{d S[\pdf]}{dt}  &=& 
- \int_{\xv} \left( \log \pdf \, \mathcal L_{\rm FP} \pdf \right) . 
\ea 
Using Eq.~(\ref{FPO-def-1}), together with the antisymmetry of $Q^{ij}$, as well as a few integrations by parts, we can show that 
\ba
\frac{d S[\pdf]}{dt}  = - \int_{\xv} \pdf \,( \nabla_i Q^{ij} )
(\nabla_j \U)  =0. 
\ea
 
Hence we find that all following properties hold about the unitary dynamics Eq.~(\ref{Langevin-f-unitary}), if $(\nabla_i  { \U} )(\nabla_j Q^{ij} )  = 0$:
\begin{enumerate}[i.]
\item The flow defined by the unitary dynamics Eq.~(\ref{Langevin-f-unitary-1}) is incompressible. 
\item The generalized potential $\U$ is conserved by the unitary dynamics Eq.~(\ref{Langevin-f-unitary-1}).
\item The Gibbs-Shannon entropy is conserved by the unitary dynamics Eq.~(\ref{Langevin-f-unitary-1}).
\end{enumerate} 
Of course, all these properties are destroyed once the noises and dissipations are turned on.

\subsection{Stochastic energetics and stochastic thermodynamics}
\label{sec:comparison-SE-ST}

The common theories of stochastic energetics and stochastic thermodynamics have two cornerstones.  Firstly, Langevin equations are formulated using the ``conventional Langevin approach''~\cite{van-Kampen-stochastic}, where one adds frictions and noises (random forces) to the otherwise unitary dynamic equations.  The resulting equations are understood as balance of forces, or, more generally, balance of thermodynamic forces, and the variances of random forces are chosen such that the steady states correspond to thermal equilibrium states in the absence of time-dependence driving forces.  Secondly, heat is defined, \`a la Sekimoto~\cite{Sekimoto-book}, as the work done to the system by the friction and random forces, and other thermodynamic variables are defined such that the first and second laws of thermodynamics are valid for arbitrary non-equilibrium processes.  

When dealing with systems whose dynamics are described by Eq.~(\ref{Langevin-f}), however, both these cornerstones are problematic.  Firstly, the Langevin equation (\ref{Langevin-f}) cannot be naively understood as condition of force balance.  In fact, if one write down force balance equation in the presence of multiplicative noises, one would obtain an equation similar to Eq.~(\ref{Langevin-f}) but with the spurious drift missing, and with detailed balance unintentionally broken.  This is in fact a well-known pathology of the ``conventional Langevin approach''~\cite{van-Kampen-stochastic,Hanggi-1980,Hanggi-1981,Grabert-1980s,van-Kampen-validity}.   Secondly, whenever spurious drift shows up, Sekimoto's definition of heat differs from ours, and leads to violation of the second law of thermodynamics.  Hence whenever the metric tensor or kinetic coefficients depend on system variables, the common theories need to be replaced by the theory developed in this work.  In Sec~\ref{sec:example} we supply three concrete examples to illustrate these points.  

It is important to emphasize that the conceptually all Langevin equations are effective dynamic equations of slow variables obtained via coarse-graining of fast variables.  This applies both to our covariant theory and to the ``conventional Langevin approach''. The difference is how the Langevin equations are formulated.  In contrast with the ad hoc way of the ``conventional Langevin approach'', in our covariant theory, Langevin equations are constructed using tensor objects, which automatically guarantees detailed balance.  



For classical point particles coupled to additive noises, our theory is in fact fully consistent with the common theories.  We first use one-dimensional system to illustrate the main point.  Following the conventional Langevin approach, we add noise and friction to the Hamiltonian equations, and obtain 
\begin{subequations}
\label{S-ud-eq}
\ba
\frac{d q}{dt} &=& \frac{p}{m}, \\
 \frac{dp}{dt} &=& - \partial_q V(q,\lambda) - \frac{ \gamma p}{m} + \eta(t),
\label{S-ud-eq-p}
\ea
\end{subequations}
where $\lambda$ is the control parameter, and  the noise correlation function is
\be
\langle \eta(t) \eta(t') \rangle = 2 \gamma T\, \delta (t-t'),
\ee
such that the system converges to an equilibrium Gibbs-Boltzmann distribution with energy $E = p^2/2m + V(q)$ and with temperature $T$.  As shown in Sec.~IV B of Ref.~\cite{covariant-Langevin-2020}, these Langevin equations can also be written in the form of covariant Langevin equations.


In the common theories, heat is defined as the work done by the friction and random force, which can be written as the following Stratonovich product: 
\ba
\dbar \mathcal Q_{\rm SE} \equiv
 \left( - \gamma \frac{d q}{dt} + \xi(t)\right) \circ dq(t),  
\label{S-def-heat}
\ea
where the subscript SE denotes Sekimoto.  The Stratonovich product $\circ$ has the special property that for any function $f(q,p)$, the usual calculus rule holds:
\be
df(q,p) = \partial_q f \circ dq + \partial_p f \circ dp.
\ee
For a detailed discussion on Stratonovich product of stochastic variables, see chapter 4 of Gardiner's book~\cite{Gardiner-book}.  Using the Langevin equation (\ref{S-ud-eq-p}), Sekimoto (see chapter 4 of Ref.~\cite{Sekimoto-book}) further showed 
\ba
\dbar \mathcal Q_{\rm SE} 
&=& \left(  \dot p + \partial_q V \right) \circ dq
\nonumber\\
&=& d \left( \frac{p^2}{2m}
 + V(q, \lambda) \right) 
- \frac{ \partial H}{\partial \lambda} d \lambda.
\ea
But this is precisely the heat defined in Eq.~(\ref{dQ-path-def-1}), with $H = p^2/2m + V$ playing the role of $H_\Xv$:
\ba
\dbar \mathcal Q &=& d H_{\Xv} - d_{\lambda} H_{\Xv}
 = d_{\xv} H_{\Xv},
\ea
where  $\xv = (q,p)$ is the shorthand for the canonical variables.  The work according to Sekimoto is then given by 
\be
\dbar \mathcal W_{\rm SE} =
 d H_{\Xv} - \dbar \mathcal Q_{\rm SE}
=  d_{\lambda} H_{\Xv}, 
\ee
which is also identical to our definitions of work, Eqs.~(\ref{dW-path-def-1}).   

}


We now consider a collection of point particles moving in a fluid and interacting with each other via certain potential $u(\rv_i -\rv_j)$.   The slow variables consist of the Cartesian coordinates $\rv_i$ and canonical momenta $\pv_i$ of all particles., i.e., $\xv = (\rv_1, \pv_1, \cdots, \rv_n, \pv_n)$.  There is also an external potential field $\psi_i(\rv, \lambda)$ acting on each particle, with $\lambda$ the external control parameter.  The total Hamiltonian is then given by
\ba
H =  \sum_i \left( \frac{\pv_i^2}{2 m_i} + \psi_i (\rv_i, \lambda)  \right) 
+ \sum_{i <j} u(\rv_i - \rv_j).  
\ea
The system is weakly coupled to a heat bath $T$, so that the equilibrium state is :
\ba
p^{\rm EQ} (\rv, \pv) = e^{\beta F - \beta H}.
\ea 
Hence the generalized potential is 
\ba
\U = \beta (H - F ).
\ea   
The nonlinear Langevin equations are given by
\begin{subequations}
\label{example-1-Langevin}
\ba
d \rv_i &=& \frac{\pv_i}{m_i} dt, \\
d \pv_i &=& - \bigg[ \sum_j \nabla_i u(\rv_i - \rv_j) 
+ \nabla_i \psi_i (\rv_i) \bigg] dt 
\nonumber\\
&&- \, \gamma_i \dot \rv_i  dt + \sqrt{2 \gamma_i T} d {\mathbf W}_i, 
\ea
\end{subequations}
where $\gamma_i$ is the friction coefficient of $i$-th particle, and $d{\mathbf W}_i$ are vector-valued Wiener noises.  Note that $\nabla_i = \partial /\partial \xv_i$ is the gradient operator in Cartesian coordinates.  Equations (\ref{example-1-Langevin}) can be written in the covariant form Eq.~(\ref{Langevin-f}), with $g = 1$, and the following non-vanishing kinetic coefficients:
\begin{subequations}
\ba
&& - Q^{r_i, p_i} = Q^{ p_i, r_i} = T, \\
&& B^{p_i, p_i} = \gamma_i T,
\ea
\end{subequations}
which satisfy  condition (\ref{DB-condition-2-L}).  There is no spurious drift since all kinetic coefficients are constants.  The remaining conditions of detailed balance are
\begin{subequations}
\ba
\psi_i (\rv_i, \lambda) &=& \psi_i (\rv_i^*, \lambda^*), \\
u(\rv) &=& u(\rv^*).  
\ea
\end{subequations}
In the limit   $\gamma_i \rightarrow 0$, Eqs.~(\ref{example-1-Langevin}) reduce to the well-known Hamiltonian equations.  

Thermodynamic quantities at ensemble level, such as system entropy, free energy etc, can be straightforwardly obtained using the results in Sec.~\ref{sec:covariant-Sto-Therm}.  Here we focus on the work and heat at trajectory level, which can be written down using Eqs.~(\ref{dW-path-def}) and (\ref{df_B-path-1}): 
\ba
\dbar {\mathcal W}  \!\! &=& d_{\lambda} H
= \sum_i ( \partial_{\lambda} \psi_i ) d \lambda,
\label{dW-ex1} \\
\dbar {\mathcal Q}  &=&  d_{\xv} H
\\
&=&  \sum_i  \left[ {-}  \frac{\gamma_i }{m_i} 
\left( \frac{\pv_i^2}{m_i} - T \right) dt
+  \sqrt{2 \gamma_i T} \,\, \dot \rv_i \cdot d {\mathbf W}_i \right] ,
\nonumber
\ea
 where we have used Ito's formula Eq.~(\ref{Ito-formula}) in the calculation of heat, and the product $\dot \rv_i \cdot d {\mathbf W}_i $ is understood in Ito's sense.  The average heat is obtained by averaging $d {\mathcal Q} $ both over noises and over $\pdf$.   The average of product $\dot \rv_i \cdot d {\mathbf W}_i $ vanishes because $\dot \rv_i $ is independent of $d {\mathbf W}_i $ in Ito-calculus.  Hence we find
\be
\dbar Q = \langle \dbar {\mathcal Q} \rangle 
=   {-} \sum_i \frac{2 \gamma_i\,dt }{m_i}  
\int_{\rv,\pv} \pdf 
\left( \frac{\pv_i^2}{2 m_i} -  \frac{3 T}{2} \right). 
\label{dQ-K-T}
\ee
 {Equations (\ref{dW-ex1}) and (\ref{dQ-K-T}) agree with those of Sec. 4.1 of Ref.~\cite{Sekimoto-book}.  Note, however, Stratonovich's calculus instead of Ito-calculus is used in Ref.~\cite{Sekimoto-book}.  }

If the process is quasi-static, which means that the system remains in equilibrium all the time, then the average of kinetic ${\pv_i^2}/{2 m_i}$ equals to $3T/2$ (the equipartition theorem), and the heat vanishes identically, and there is no dissipation of energy.   On the other hand, Eq.~(\ref{dQ-K-T}) also means that as long as there is energy dissipation due to friction, the average kinetic energy cannot satisfy the equipartition theorem.  This implies that the over-damped limit of a Hamiltonian cannot be obtained by simply assuming that the momentum remains in equilibrium, for that would imply no friction.  

Using the covariant properties established in this work, we are able to carry out arbitrary nonlinear transformation of variables, which can be either canonical or non-canonical.  This can be very useful when dealing with multi-dimensional Hamiltonian systems with various kind of nonlinear constraints.  More examples will be explored in future works. 

{
\section{Three Examples}
\label{sec:example}

In this section, we discuss three examples where the common theories fail due to the presence of spurious drift, whereas our theory yields correct results.  In the first example, Langevin equations are formulated in curvilinear coordinates, whereas in the second example, the slow manifold is curved.  In the third example, the kinetic coefficient is state-dependent.  More results will be supplied in the future. 



\subsection{Diffusion in polar coordinates}
We consider the over-damped limit of two dimensional Brownian motion, subjected to a potential  $V(r; \lambda)$ with rotational symmetry and/or with radial boundary condition $r < r_0$.  It is then more convenient to use polar coordinates.  We first start from the over-damped Langevin equations in Cartesian coordinates:
\begin{subequations}
\label{Langevin-polar-0}
\ba
\gamma \, dx + \partial_x V(r; \lambda) \,dt &=& \sqrt{2 T \gamma} \, d W_x, \\
\gamma \, dy + \partial_y V(r; \lambda)  \, dt&=& \sqrt{2 T \gamma} \, d W_y,
\ea
\end{subequations}
which can be readily understood as conditions of force balance.  The noise amplitude $\sqrt{2 T \gamma}$ is such that the steady state distribution is the thermal equilibrium state $e^{-\beta V(r)}dx dy$.   Equations (\ref{Langevin-polar-0}) can be written into the matrix form 
\ba
\begin{pmatrix}
 dx \\  dy
\end{pmatrix}
+ \begin{pmatrix}
T/\gamma & 0 \\
0 & T/\gamma  \end{pmatrix}
\begin{pmatrix}
\partial_x \beta V \\ 
\partial_y \beta V
\end{pmatrix}
= \begin{pmatrix}
\sqrt{2 T /\gamma} \, dW_x 
\vspace{1mm}\\ 
\sqrt{2 T /\gamma}\, dW_y
\end{pmatrix}, 
\nonumber\\
\ea
which is our standard form, Eq.~(\ref{Langevin-f}), with $\U = \beta H_\Xv = \beta V, g = 1$, and 
\ba
\Lm =  \frac{T}{\gamma} \begin{pmatrix}
1& 0 \\
0& 1
\end{pmatrix}, \quad
{\boldsymbol b} =  \sqrt{\frac{2 T}{\gamma}} \begin{pmatrix}
1 & 0 \\ 0 & 1 \end{pmatrix}.
\ea
Note that $\Lm $ is a constant matrix, hence there is no spurious drift. 

There are two possible ways to obtain Langevin equations in polar coordiates: (i) We may use the transformation rules Eqs.~(\ref{transform-U-L-b-new}) to obtain various tensors in polar coordinates and then assemble the Langevin equation in polar coordinates using  Eq.~(\ref{Langevin-f}).  (ii) Alternatively we may also apply NTV and Ito's formula (\ref{Ito-formula}) directly to Eqs.~(\ref{Langevin-polar-0}), and obtain the new equations.  The results are of course the same.  Here we take the first route. 

The transformation matrix from Cartesian coordinates to polar coordinates is
\ba
\frac{\partial (r, \phi)}{\partial(x, y)}
= \begin{pmatrix}
\cos \phi & \sin \phi \\
- \sin \phi /r & \cos \phi /r
\end{pmatrix}.  
\ea
Using Eqs.~(\ref{transform-U-L-b-new}), we obtain the transformed kinetic matrix and noise amplitude:
\ba
\Lm' =  \frac{T}{\gamma} \begin{pmatrix}
1& 0 \\
0& 1/ r^2
\end{pmatrix}, \quad
\boldsymbol b' =   \sqrt{\frac{2 T}{\gamma}}
\begin{pmatrix}
\cos \phi & \sin \phi \\
- \sin \phi /r & \cos \phi /r
\end{pmatrix}.  
\nonumber\\
\ea
The determinant of the new metric tensor is $g' = r^2$.  The generalized potential $\U$ remains the same since it is a scalar.   Using these results and further defining new Wiener noises
\begin{subequations}
\ba
dW_r &=& \cos \phi \, dW_x + \sin \phi \, dW_y, \\
dW_r &=& -\sin \phi   \, dW_x + \cos \phi \, dW_y, 
\ea
\end{subequations}
we obtain the transformed Langevin equations:
\begin{subequations}
 \label{polar-dr-dtheta-1}
 \ba
dr + \frac{1}{\gamma} \left( 
\partial_r V - \frac{T}{r} \right) dt
 &=&  \sqrt{\frac{2 T}{\gamma}} dW_r, 
 \label{polar-dr-1}\\
d \phi  &=&  \sqrt{\frac{2 T}{\gamma}} \frac{1}{r} dW_\phi.  
\ea 
\end{subequations}
Note that the term $- {T\, dt}/{\gamma\,r} $ in Eq.~(\ref{polar-dr-1}) is the spurious drift, resulting from the $r$ dependence of the metric.  The associated Fokker-Planck equation is:
\be
\partial_t \pdf = \frac{T}{\gamma r} \partial_r 
r \left( \partial_r \pdf  + (\partial_r \beta V) \pdf \right)
+ \frac{T}{\gamma r^2} \partial_\phi^2 \pdf. 
\label{FPE-polar}
\ee 
The heat and work at trajectory level are
\begin{subequations}
\ba
d \mathcal Q  &=& T d_r \U 
=  d_r V = (\partial_r V) dr + \frac{T\,(\partial_r^2 V)}{\gamma } dt,
\quad \label{dQ-polar-1}  \quad \\
d \mathcal W  &=& T d_\lambda \U  
= d_\lambda V = (\partial_\lambda V) d\lambda. 
\ea
\end{subequations}
The total entropy production rate is given by Eq.~(\ref{dS^tot-2}).  Using the above results, we obtain
\ba
\frac{dS^{\rm tot}}{dt} =  \int_{r, \phi} \frac{T} {\gamma \, \pdf} 
 \left[ (\partial_r \pdf + (\partial_r \beta V) \pdf)^2
 + \frac{(\partial_\phi \pdf)^2}{r^2}  \right], 
 \quad
\ea
which is non-negative, and vanishes identically at equilibrium where $\pdf \sim e^{-\beta V}$.   

If one try to write down Langevin equations in polar coordinates using the idea of force balance, one would obtain 
\begin{subequations}
\label{polar-dr-dtheta-2}
\ba
-V' - \gamma \dot r 
+  \sqrt{{2 T \gamma}} \frac{dW_r}{dt} &=&  0,
\label{polar-dr-2} \\
 - \gamma r\,\dot \phi  + \sqrt{{2 T \gamma}} \frac{dW_\phi}{dt} &=& 0.
\ea
\end{subequations}
which, comparing with the correct equations (\ref{polar-dr-dtheta-1}), misses the spurious drift.  It is important to note that noises are additive in this problem, hence there is no issue of stochastic calculus here. 

Additionally, if we adopt Sekimoto's definition of heat as the work done by friction and random force, and use the correct Langevin equations (\ref{polar-dr-dtheta-1}), we would have
\ba
d\mathcal Q_{\rm SE} &=&  \left(- \gamma \dot r 
+ \sqrt{{2 T \gamma}} \frac{dW_r}{dt} \right) 
\circ dr 
\nonumber\\
&+&  \left( - \gamma r\,\dot \phi  + \sqrt{{2 T \gamma}} \frac{dW_\phi}{dt} \right)
\circ r d \phi  
\nonumber\\
&=&  \left( V'(r) - T/r \right) \circ dr 
\nonumber\\
&=& d \mathcal Q - T \, d \log r , 
\label{dQ-SE-1}
\ea 
which differs from  Eq.~(\ref{dQ-polar-1}) by $-T d \log r$.  Using this to calculate the total entropy production, we would obtain 
\ba
\frac{dS^{\rm tot}_{\rm SE}}{dt} &=&
\frac{dS}{dt}
- \beta \left\langle \frac{d \mathcal Q_{\rm SE}}{dt} \right\rangle
 \nonumber\\
 &=& \frac{dS^{\rm tot}}{dt} + \left\langle d_r \log r \right\rangle
 \nonumber\\
 &=&  \frac{dS^{\rm tot}}{dt} 
 - \frac{1}{\gamma} \int_{r, \phi} \frac{\partial_r V}{r} \pdf,  
 \label{dS-tot-polar-SE}
\ea 
where $ \left\langle \,\,  \cdot \,\, \right\rangle$ means average over both probability distribution and noises, which we have calculated using Eqs.~(\ref{polar-dr-2}) and Ito formula $dr dr = 2 T dt/\gamma$ (obtained from Eq.~(\ref{Ito-formula})).     At thermal equilibrium, $\pdf^{\rm EQ}(r) \propto e^{-\beta V(r)}$, and the extra term in Eq.~(\ref{dS-tot-polar-SE}) can be calculated:
\ba
 - \frac{1}{\gamma} \int_{r, \phi} \frac{\partial_r V}{r} \pdf
 = \frac{2 \pi T}{\gamma} \left( \pdf^{\rm EQ}(r_0) - \pdf^{\rm EQ}(0) \right),  
\ea
which is non-vanishing  as long as $V(r_0) \neq V(0)$.  Since ${dS^{\rm tot}}/{dt} = 0$ at equilibrium, Eq.~(\ref{dS-tot-polar-SE}) predicts a non-vanishing entropy production rate at equilibrium for $V(r_0) \neq V(0)$, which violates the second law of thermodynamics.  

It is important to note that for this particular simple problem, it is possible to interpret the spurious drift as an additional friction force due to geometry of curvilinear coordinates, such that Sekimoto's definition of heat becomes correct.  Such a procedure however does not work for general multi-dimensional problems, because the spurious drift $(1/\sqrt{g}\, \partial_j (\sqrt{g} L^{ij})$ involves both reactive and dissipative parts of $L^{ij}$ and cannot not be interpreted as friction force.  }

\subsection{Diffusion on a sphere}
\label{sec:sphere} 
As a similar problem, we consider diffusion of a molecule on a unit sphere, or the rotational diffusion of an uniaxial particle, subjected to an external potential $V(\theta; \lambda)$ that is invariant under rotation around $z$ axis.   We shall again consider the {\em over-damped limit}, hence the slow variables are the polar and azimuthal angles $\theta, \phi$.  Unlike the preceding example, however, we have no Cartesian coordinates to start with, but have to construct the equations in polar coordinates directly.     The metric  has the following form:
\be
ds^2 = g_{\theta\theta} d\theta^2 + g_{\phi\phi} d\phi^2
=  d \theta ^2 + \sin^2 \theta d \phi^2 . 
\ee
 The covariant and contra-variant  metric tensors ${\mathbf g}$ then have the following components:
\ba
&& g_{\theta\theta} = 1, \quad g_{\phi\phi} = \sin^2 \theta, 
\quad g_{\theta\phi} = g_{\phi\theta} = 0, \\
&& g^{\theta\theta} = 1, \,\, g^{\phi\phi} = \sin^{-2} \theta , 
\,\, g^{\theta\phi} = g^{\phi\theta} = 0.  \quad\quad
\ea
The corresponding matrix forms are ${\mathbf g}$ and ${\mathbf g}^{-1}$, given by:
\be
{\mathbf g} = \begin{pmatrix}
1 & 0\\
0 &  \sin^{2} \theta 
\end{pmatrix}, \quad
{\mathbf g}^{-1} = \begin{pmatrix}
1 & 0\\
0 &  \sin^{-2} \theta 
\end{pmatrix}.
\ee
The invariant volume measure is then 
\be
dv = \sqrt{g} d \theta d \phi = \sin \theta d \theta d \phi. 
\ee

We shall assume that there is an external potential which only depends on the polar angle $\theta$, so that  the equilibrium pdf is $e^{-\beta V(\theta, \lambda)}\sin \theta d \theta d \phi$.  Since we are studying the over-damped theory, detailed balance demands that the antisymmetric matrix $Q^{ij}$ vanishes identically.  We shall assume that the symmetric matrix $B^{ij}$  has rotational symmetry.  It is known that $g_{ij}$ and $g^{ij}$ are the only covariant and contra-variant tensors on the sphere with rotational symmetry.  Hence we see that $\Bm$ must be proportional to ${\mathbf g}^{-1}$:
\be
\Bm = \begin{pmatrix}
D & 0\\
0 & D  \sin^{-2} \theta
\end{pmatrix},
\ee
where $D$ is the diffusion constant.  

The Langevin equations can be obtained from Eq.~(\ref{Langevin-f}) via proper choices of the matrix $b^{i \alpha}$: 
\begin{subequations}
\label{Langevin-sphere}
\ba
d \theta + D \left( \beta \partial_\theta V -  \cot \theta \right) dt 
&=& {\sqrt{2D}} \, dW_{\theta}, 
\quad\quad
\label{Langevin-sphere-a}\\
d \phi &=& \frac{\sqrt{ 2D}}{ \sin \theta} \, d W_{\phi},
\ea
where the term $- D\, \cot \theta$ in Eq.~(\ref{Langevin-sphere-a}) is spurious drift. 
\end{subequations}
The associated Fokker-Planck equation is:
\be
\partial_t \pdf = \frac{D}{\sin \theta} \partial_\theta 
\sin \theta \left( \partial_\theta \pdf  + \beta (\partial_\theta V) \pdf \right)
+ \frac{D}{\sin^2 \theta} \partial_\phi^2 \pdf. 
\label{FPE-sphere}
\ee 
 {Equations (\ref{Langevin-sphere}) and (\ref{FPE-sphere}) agree with the results obtained by Raible and Engel~\cite{Raible2004}.   }  They also reduce to Eqs.~(\ref{polar-dr-dtheta-1}) and (\ref{FPE-polar}) as $\sin \theta \sim r \rightarrow 0$ with $D = T/\gamma$.   Using  Eqs.~(\ref{dW-path-def}), (\ref{df_B-path-1}) and (\ref{U-F-B}), the differential heat and work at trajectory level are
\ba
\dbar { \mathcal Q} &=&  (\partial_\theta V ) d \theta 
+ D\, (\partial_\theta^2 V ) dt,\\
\dbar W &=&  (\partial_\lambda V  ) d\lambda.
\ea
{ The total entropy production rate can again be calculated using Eq.~(\ref{dS^tot-2}): 
\ba
\frac{dS^{\rm tot}}{dt} =  \int_{\theta, \phi} \frac{D} {\pdf} 
 \left[ (\partial_\theta \pdf + (\partial_\theta \beta V) \pdf)^2
 + \frac{(\partial_\phi \pdf)^2}{\sin^2 \theta}  \right], 
 \quad
\ea
which is non-negative, and vanishes identically at equilibrium where $\pdf \sim e^{-\beta V(\theta; \lambda) }$.   

Similar to the first example, if one follow the conventional Langevin approach, incorrect Langevin equations is obtained.  If one use Sekimoto's definition of heat, incorrect result for total entropy production is obtained, and the second law is violated.  The derivations are very similar, so we skip them here. 
}


{ 
\subsection{Diffusion with position-dependent friction}  
\label{sec:variant-gamma}
Consider diffusion of a particle in heterogeneous environment with position-dependent friction coefficient $\gamma (x)$ and in an external potential $V(x)$.  \xing{(Note this is not the same problem as studied in Refs.~\cite{Bo2017,Bo2019,Celani2012}, where temperature varies over space.)} The underdamped Langevin equations are
\begin{subequations}
\ba
 dx &=& \frac{p}{m}dt,
\\ 
d p &=& -\frac{ \gamma(x) p} {m} dt- V'(x) dt 
+\sqrt{2 T \gamma(x)} dW,
\quad\quad
\label{underdamped}
\ea
\label{underdamped-full}
\end{subequations}
where $dW$ is the Wiener noise and $V'(x)$ is spatial derivative of the potential.   These equations can be rewritten as the covariant form Eq.~(\ref{Langevin-f}) with
\ba
&& \U(x,p) = \beta ( H -  F )
= \beta \left( \frac{ p^2}{ 2m} + V(x) -F \right), 
\quad\quad\label{underdampedU}
\\
&& Q = \begin{pmatrix} 
0 & -T \\ T & 0
\end{pmatrix}, 
\quad
 B = \begin{pmatrix} 
0 & 0 \\ 0 & \gamma(x) T
\end{pmatrix},
\ea
where the first columns (rows) of the matrices refer to $x$ and the second columns (rows) refer to $p$. There is no spurious drift because $\gamma(x)$ depends only on $x$ but not on $p$.   Detailed balance is clearly satisfied.  

We are interested in the over-damped limit of this problem, where $\gamma(x)$ is very large (or equivalently, the mass is very small), so that only the dynamics of position becomes relevant.  In Ref.~\cite{Durang2015}, it was shown through a very long calculation that the over-damped equation is
\ba
d x + \frac{V'(x)}{\gamma(x)} dt 
- \left(\frac{T}{\gamma(x)} \right)' dt 
= \sqrt{\frac{ 2 T}{\gamma(x)}} dW,
\label{correct-over-damped}
\ea
where the third term in LHS  is the spurious drift.  The over-damped equation can also be obtained using multi-scale perturbation method.  The calculation is however still rather complicated.  We note that if the friction is independent of position, Eq.~(\ref{correct-over-damped}) reduces to the well-known result:
\be
d x +\frac{V'(x)}{\gamma} dt = \sqrt{\frac{ 2 T}{\gamma}}  dW. 
\label{correct-over-damped-0}
\ee

Using our covariant formalism of Langevin dynamics, however, the over-damped equation (\ref{correct-over-damped}) can be obtained almost instantaneously.   The covariant Langevin equation, Eq.~(\ref{Langevin-f}), tells us that  the over-damped equation must have the following form:
\ba
dx + B(x) \U_{od}'(x)dt - B'(x) dt= \sqrt{2 B(x)} dW.
\label{correct-over-damped-B}
\ea
Recall that the equilibrium pdf of the over-damped theory is $e^{- \U_{od}(x)}$, and obtained form that of the under-damped theory by integrating out $p$:
\ba
e^{- \U_{od}(x)} = \int  e^{ - \U(x,p)} dp,
\ea
where $ \U(x,p)$ is given in Eq.~(\ref{underdampedU}).  Hence we find
 \ba
 \U_{od} (x) = \beta V(x)  + C.
 \ea 
  Now Eq.~(\ref{correct-over-damped-B}) must reduce to Eq.~(\ref{correct-over-damped-0}) if the friction is independent of $x$.  Hence $B(x) = T/ \gamma(x)$, and Eq.~(\ref{correct-over-damped-B}) is indeed identical to Eq.~(\ref{correct-over-damped}).
This example illustrates the power of our covariant Langevin dynamics. 
     
Using Eq.~(\ref{df_B-path-1})  the heat at trajectory level is
\ba
\dbar { \mathcal Q}
= d_x V(x) = V' \circ dx
= V' dx + \frac{T}{2\gamma} V'' dt , \quad
\label{correct-heat}
\ea
where $\circ$ means product in Stratonovich's sense.   The total entropy production rate is calculated using Eq.~(\ref{dS^tot-2}):
\ba
\frac{dS^{\rm tot}}{dt} &=& dS - \beta \dbar \mathcal Q
\nonumber\\
&=& \int_x \frac{T}{\gamma \pdf} 
(\partial_x \pdf + (\partial_x \beta V) \pdf)^2,
\ea
which is non-negative and vanishes at the equilibrium where $\pdf \sim e^{-\beta V}$.  

If we follow the conventional Langevin approach to write down an over-damped equation based on force balance, we would obtain
\ba
\gamma(x) d x + V'(x)dt =  \sqrt{2 T \gamma(x)} \, dW,
\label{wrong-over-damped}
\ea
which misses the spurious drift completely. 
The heat defined by Sekimoto is
\ba
\dbar { \mathcal Q}_{\rm SE} &=&
 \left ( -\gamma dx/dt + \sqrt{2 T \gamma(x)} dW/dt \right) \circ dx 
\nonumber\\
&=& \left( V' + T \gamma'/\gamma \right) \circ dx
\nonumber\\
&=& d_x V(x) + T \, d_x  \log \gamma(x)
\nonumber\\
&=& \dbar { \mathcal Q} + T \, d_x  \log \gamma(x),
\label{wrong-heat}
\ea
where in the second equality we have used the correct Langevin equation (\ref{correct-over-damped}).  This definition of heat differs from the correct definition (\ref{correct-heat}) by a term 
\ba
T \, d_x  \log \gamma(x) = T \frac{\gamma'}{\gamma} dx
+ \frac{T^2}{2\gamma}\left( \frac{\gamma'}{\gamma} \right)'dt,
\label{extra-term}
\ea  
where we have used Ito's formula (\ref{Ito-formula}) to expand $d_x  \log \gamma(x)$ to the second order in $dx$.  

For simplicity, let us assume that there is no the external potential, i.e., $V = 0$, and the system is confined in an interval with length $L$.  The equilibrium state then corresponds to a flat distribution $\pdf^{\rm EQ}(x) =1/L$.   To calculate the ensemble average of Eq.~(\ref{extra-term}), we use Eq.~(\ref{correct-over-damped}) to replace $dx$, average over noise, and  multiply by $\pdf^{\rm EQ}(x)$, and finally integrate over $x$, and obtain:
\ba
\left\langle T \, d_x  \log \gamma(x) \right \rangle_{\rm EQ}
&=& \left. \frac{dt \, T^2}{L} \left(\frac{\gamma'}{\gamma^2} \right) \right|^L_0,
\ea
which is generically non-vanishing.  Hence, using of Eq.~(\ref{wrong-heat}) as definition of heat leads to a non-vanishing total entropy production rate in the equilibrium state:
\ba
\frac{dS^{\rm tot}_{\rm SE}}{dt}  =
-  \left. \frac{dt \, T}{L} 
\left(\frac{\gamma'}{\gamma^2} \right) \right|^L_0,
\ea
which violates the second law of thermodynamics.   
}






\section{Conclusion}
\label{sec:conclusion}
In this work, we have developed a covariant  theory of thermodynamics and stochastic thermodynamics for non-equilibrium small systems.  As we have demonstrated using concrete examples, this theory should be used in replacement of the common theories of stochastic energetics and stochastic thermodynamics if the noises are multiplicative, the slow variable space is curved, or curvilinear coordinates are used.  \xing{In future, we will use the theory to study more sophisticated models of Langevin dynamics with spurious drift.  
One interesting possibility is Brownian motion of rigid body and the associated theory of stochastic thermodynamics.  We will also study small systems that are driven by non-conservative forces, or with temperature gradients.  We will also study how the theory transform under coarse-graining, thereby establish the connections between different levels of stochastic thermodynamics, and understand how entropy productions at different levels of theory are related to each other.   Finally we shall also try to generalize the theory to quantum systems.  }

\section{Acknowledgement}
The research is supported by NSFC grant 11674217 and Shanghai Municipal Science \& Technology Major Project (Grant No.2019SHZDZX01).  X.X. is also thankful to additional support from a Shanghai Talent Program.
 


\appendix

\begin{widetext} 

\section{Alternative Proof of  Eq.~(\ref{DB-condition-3-new})}
\label{sec:app}
In this appendix, we explicitly calculate the short time transition probability $dv(\xv_1) \pdf_F (\xv_1,t+dt | \xv;  t )$ using the Langevin equation (\ref{Langevin-f}), and prove the formula Eq.~(\ref{DB-condition-3-new}).  

\vspace{-5mm}
\subsection{Short-time transition probability}
\label{sec:short-time-transition}
Let us for now consider the following general form of Ito-Langevin equation:
\be
dx^i - F^i(\xv,\lambda ) dt =  b^{i \alpha} (\xv, \lambda )  dW_{\alpha},
\label{Langevin-general}
\ee
{where $F^i(\xv,\lambda)$ is usually called the {\em systematic force}.}  Note that both $F^i(\xv,\lambda) $ and $b^{i \alpha} (\xv, \lambda )$ may depend on the control parameter $\lambda$ which may vary with time.    Consider a transition from $\xv$ at time $t$ to  $\xv_1$ at time $t + dt$, and let $\alpha \in (0,1)$, so that $\xv_\alpha = \xv + \alpha (\xv_1 - \xv) =  \xv + \alpha d \xv$ is an intermediate point between $\xv$ and $\xv_1$.  In Ref.~\cite{covariance-path-integral-2021} (Eq.~(68), with $\bar \alpha = 0$, or equivalently, Eqs.~(37)), we proved that the differential transition probability can be written as:
\begin{subequations}
\ba
dv(\xv_1) \pdf_F\xv_1,t+dt | \xv;  t ) &=& 
\frac{d \mu(\xv_1) \, e^{- A^\alpha (\xv_1, \xv; d t, \lambda )}}
{\sqrt{(4 \pi d t)^{n} \det B^{ij}(\xv_{\alpha}, \lambda )} } ,
\label{tran-prob-3} 
\\ \nonumber 
A^\alpha(\xv_1 , \xv; dt, \lambda ) &=&  \bigg[  d x^i 
-  d t \left( F^i   - 2 \alpha \partial_k B^{ik}  \right)_{\! \alpha} \bigg]
\frac{B^{-1}_{ij}(\xv_{\alpha}, \lambda ) }{4d t} 
 \bigg[ d x^j  - d t  \left( F^j 
  - 2 \alpha \partial_l B^{jl}  \right)_{\alpha} \bigg] 
 \\ 
 &+ & \alpha \,  ( \partial_i F^i )_{ \alpha} d t 
  - \alpha^2 ( \partial_i \partial_j B^{ij} )_{\alpha} d t,  
  \label{tran-prob-alpha}
\ea
 \end{subequations}
where $d\mu(\xv_1) = d^n\xv_1$ is an infinitesimal volume element around $\xv_1$, and $d v(\xv_1) = \sqrt{g(\xv_1)}d^n \xv_1$ is the invariant volume measure, $B^{-1}_{ij}$ is the inverse matrix of $B^{ij} = b^{i \alpha} b^{j \alpha}$, see Eq.~(\ref{B-b-relation}),  {and $(\cdots)_\alpha$ means that  all functions inside the bracket are evaluated at the intermediate point $\xv_\alpha$.   The matrix $B^{ij}$  is  assumed to satisfy the detailed balance condition Eq.~(\ref{DB-condition-2-B}).   }

In particular, the $\alpha=0$ version of short-time transition probability is:
\ba
dv(\xv_1) \pdf_F\xv_1,t+dt | \xv;  t ) &=& 
\frac{d \mu(\xv_1) \, e^{- A^0 (\xv_1, \xv; d t, \lambda )}}
{\sqrt{(4 \pi d t)^{n} \det B ^{ij}(\xv ,\lambda)} } ,
 \label{tran-prob-1}
\\ 
A^0(\xv_1 , \xv; d t, \lambda ) &=&  \left(  d x^i - F^i(\xv, \lambda) d t \right)
\frac{B^{-1}_{ij}(\xv,\lambda ) }{4 d t} 
 \left( d x^j  - F^j (\xv, \lambda) d t  \right) .
 \label{action-1}
\ea
Note that the linear combination $dx^i - dt F^i ( \xv,t)$ in the action Eq.~(\ref{action-1})  is precisely the deterministic part of the Langevin equation (\ref{Langevin-general}).  In our covariant formulation of Langevin theory, this linear combination transforms as a contra-variant vector under under nonlinear transformation of variables, even though neither $dx^i$ or $F^i $ does so. 


The defining features of Eq.~(\ref{tran-prob-1}) are that $d\xv$ is  Gaussian random variable, and has the following moments (here $\kappa_m$ denotes cumulant of order $m$):
\begin{subequations}
\label{moments-dx}
\ba
&& \langle { dx}^i \rangle = F^i  (\xv, \lambda) dt, \\
 &&  \langle { dx}^i { dx}^i \rangle
  -    \langle { dx}^i \rangle \langle { dx}^j \rangle
 = 2 B^{ij} (\xv, \lambda) dt, \\
&& \kappa_m ( { d\xv} ) = 0, \quad \forall \,\,m \geq 3. 
\ea
\end{subequations}

For $\alpha \neq 0$, the action  Eq.~(\ref{tran-prob-alpha}) is not quadratic in $d\xv$.   Hence the transition probability Eq.~(\ref{tran-prob-3}) is not Gaussian in $d\xv $.  However, in Ref.~\cite{covariance-path-integral-2021}, we proved that all moments of ${d\xv} $ obtained from Eq.~(\ref{tran-prob-3}) are independent of $\alpha$ up to order $dt$.  This means that in the short-time limit, $dt \rightarrow 0$, Eq.~(\ref{tran-prob-3}) with different $\alpha$ are equivalent to each other, in the sense that they all generate the same continuous-time Markov process, with the same statistical properties of physical observables.   In another word, the non-Gaussian nature of Eq.~(\ref{tran-prob-3}) for $\alpha \neq 0$ makes no contribution to the statistical properties of the Markov process in the continuous time limit.  

\subsection{Reversible and Irreversible Forces}
The systematic force $F^i  (\xv, \lambda)$ in Eq.~(\ref{Langevin-general}) can be decomposed into a reversible part $ F^i _{(R)} (\xv, \lambda)$ and an irreversible part $F^i _{(I \! R)}  (\xv, \lambda)$, which are respectively defined as
\ba
 F^i _{(R)} (\xv, \lambda)  &\equiv& 
 \frac{1}{2} \left( 
 F^i (\xv, \lambda) - \epsilon_i  F^i (\xv^*, \lambda^*) \right),
  \\
 F^i _{(I \! R)} (\xv, \lambda)  &\equiv& 
 \frac{1}{2} \left( 
 F^i (\xv, \lambda) + \epsilon_i  F^i (\xv^*, \lambda^*) \right),
\\
F^i  (\xv, \lambda) &=&  F^i _{(I \! R)}(\xv, \lambda) + F^i _{(R)}(\xv, \lambda).
\label{F-decomp-1} 
\ea
{ Here and below, we shall hide the time-dependence of $F^i $ and $B^{ij}$, to simplify the notations.  } It then follows from these definitions that (no summation over repeated indices below)
\begin{subequations}
\label{F-decomp-app}
\ba
\epsilon_i F^i _{(R)} (\xv^*, \lambda^*) &=& - F^i _{(R)}(\xv, \lambda)  , \\
\epsilon_i F^i _{(I \! R)} (\xv^*, \lambda^*) &=& F^i _{(I \! R)}(\xv, \lambda) , \\
\epsilon_i  F^i  (\xv^*, \lambda^*) &=&  F^i _{(I \! R)}(\xv, \lambda)  - F^i _{(R)}(\xv, \lambda)  . 
\ea
\end{subequations}
 
For the covariant Langevin equation (\ref{Langevin-f}), we have 
\begin{subequations}
\label{F-decomp-def}
\ba
F^i (\xv, \lambda)  &=& \partial_j L^{ij}(\xv, \lambda) 
 - L^{ij}(\xv, \lambda) \, \partial_j  \!
 \left[ \U(\xv, \lambda) - \log \sqrt{g(\xv)}\right]. 
\ea
Using the conditions of detailed balance, Eqs.~(\ref{DB-condition-2}), we can show that the reversible and irreversible parts of the systematic force are 
\ba
F^i _{(R)}(\xv, \lambda)  &=& \partial_j Q^{ij} (\xv, \lambda)
   - Q^{ij}(\xv, \lambda)  \, \partial_j \! \left[ \U(\xv, \lambda) - \log \sqrt{g(\xv)}\right],
   \label{F-decomp-def-3}\\
    F^i _{(I \! R)}(\xv, \lambda)  &=& \partial_j B^{ij}(\xv, \lambda)  
 - B^{ij} (\xv, \lambda)  \, \partial_j \! \left[ \U(\xv, \lambda) - \log \sqrt{g(\xv)}\right], 
    \label{F-decomp-def-4}\\
F^i (\xv, \lambda)  &=& 
F^i _{(I \! R)}(\xv, \lambda)  + F^i _{(R)} (\xv, \lambda). 
\ea 
\end{subequations}
Hence the reversible force only involves $Q^{ij}$ whereas the irreversible force only involves $B^{ij}$.  This is, of course, consistent with our understanding that $Q^{ij}$ are the reactive couplings whereas $B^{ij}$ are the dissipative couplings.  

\subsection{Proof of Eq.~(\ref{DB-condition-3-new})}

 {To calculate the  ratio (\ref{DB-condition-3-new}), it is most convenient to use $\alpha = 1/2$ version of the transition probability (\ref{tran-prob-3}): }
\ba
dv(\xv_1) \pdf_F\xv_1,t+dt | \xv;  t ) &=& 
\frac{d \mu(\xv_1) \, e^{- A^{1/2} (\xv_1, \xv; d t, \lambda )}}
{\sqrt{(4 \pi d t)^{n} \det B^{ij}(\xv_{1/2} )} } ,
\label{tran-prob-3-app} 
\\ \nonumber 
A^{1/2}(\xv_1 , \xv; dt, \lambda ) &=&  \bigg[  d x^i 
-  d t \left( F^i   -  \partial_k B^{ik}  \right)_{\! 1/2} \bigg]
\frac{B^{-1}_{ij}(\xv_{1/2}) }{4d t} 
 \bigg[ d x^j  - d t  \left( F^j 
  -  \partial_l B^{jl}  \right)_{1/2} \bigg] 
 \\ 
 &+ & \frac{1}{2} \,  ( \partial_i F^i )_{1/2} d t 
  - \frac{1}{4} ( \partial_i \partial_j B^{ij} )_{1/2} d t,  
 \label{action-forward}
\ea
where $F^i, B^{ij}$ depend both on $\xv_{1/2}$ and on $\lambda$.  Using Eq.~(\ref{F-decomp-1}) and introduce a shorthand $\yv = \xv_{1/2} = (\xv_1 + \xv)/2$, we can rewrite Eq.~(\ref{action-forward}) as
 \ba
A^{1/2} (\xv_1, \xv, dt, \lambda)  &=&
  ( dx^i - dt ( F^i _{(I \! R)}(\yv, \lambda)  + F^i _{(R)}(\yv, \lambda)  
 -  \partial_k  B^{ik}(\yv , \lambda) ) ) 
  \nonumber\\
\times  \frac{ B^{-1}_{ij}  (\yv, \lambda) }{4 \,dt} 
 &\times&  ( dx^j - dt ( F^j_{(I \! R)}(\yv, \lambda) + F^j_{(R)}(\yv, \lambda)  
  -  \partial_l   B^{jl}(\yv, \lambda)  ) )
\nonumber\\
&+& \frac{1}{2} dt \, \partial_i ( F^i _{(I \! R)}(\yv, \lambda) 
 + F^i _{(R)}(\yv, \lambda)  )
 -  \frac{1}{4}  dt \, \partial_i \partial_j  B^{ij}(\yv, \lambda) .  
 \label{action-forward-1}
\ea

Now consider the backward process where the system goes from $\xv_1^*$ to $\xv^*$, and the parameter is fixed at $\lambda^*$.  Note that the mid-point of the backward process is $\yv^* = \xv^* + d\xv^*/2$.  Let us write down the $\alpha = 1/2$ version of the transition probability for the backward process, which is the counterpart of Eq.~(\ref{tran-prob-3-app}):
\ba
d v(\xv^*)  \pdf_B (\xv^* , - t  | \xv_1^* , -t - dt) 
&=& 
\frac{  e^{- A^{1/2}( \xv^* , \xv_1^*, dt, \lambda^*) } d\mu( \xv^*)}
{\sqrt{(4 \pi dt)^n \det B^{ij}( \yv^*, \lambda^*)}},
\label{tran-prob-back} \\
A^{1/2} (\xv^* , \xv_1^*, dt, \lambda^*) &=&
  ( - (dx^i)^* - dt ( F^i  (\yv^*, \lambda^* )  
 -  \partial_k^* B^{ik}(\yv^*, \lambda^* ) ) ) 
  \nonumber\\
\times  \frac{ B^{-1}_{ij}  (\yv^*, \lambda^*)}{4 \,dt} 
 &\times&  (- (dx^j)^* - dt ( F^j (\yv^*, \lambda^* ) 
  -  \partial_l^* B^{jl}(\yv^*, \lambda^*)  ) )
\nonumber\\
&+& \frac{1}{2} dt  \,  \partial_i^* F^i  (\yv^*, \lambda^*)
 -  \frac{1}{4}  dt  \, \partial_i^* \partial_j^* B^{ij}(\yv^*, \lambda^*).  
 \label{action-back}
\ea
Using $(x^i)^* = \epsilon_i x^i$, we can rewrite the action Eq.~(\ref{action-back}) into
\ba
A^{1/2} (\xv^*, \xv_1^*, dt ,\lambda^*)   &=&
  ( - dx^i - dt ( \epsilon_i F^i  (\yv^*, \lambda^* )  
 -  \partial_k  \epsilon_i   B^{ik}(\yv^*, \lambda^* )  \epsilon_k  ) ) 
  \nonumber\\
 \times  \frac{ \epsilon_i  B^{-1}_{ij}  (\yv^*, \lambda^*) \epsilon_j }{4 \,dt} 
 &\times&  ( - dx^j - dt (  \epsilon_j  F^j (\yv^*, \lambda^* )  
  -  \partial_l  \epsilon_j  B^{jl}(\yv^*, \lambda^*) \epsilon_l  ) )
\nonumber\\
&+& \frac{1}{2} dt \, \partial_i  \epsilon_i  F^i  (\yv^*, \lambda^*)
 -  \frac{1}{4}  dt \, \partial_i \partial_j  \epsilon_i  B^{ij}(\yv^*, \lambda^*)  \epsilon_j .  
 \label{action-back-1}
\ea 
{Using Eq.~(\ref{F-decomp-1}) to decompose $F^i  $ into $F_{(IR)}^i + F_{(R)}^i$, and further using the symmetry properties, Eqs.~(\ref{F-decomp-app}), as well as the conditions of detailed balance (\ref{DB-condition-2-B}), we can rewrite Eq.~(\ref{action-back-1}) into  }
\ba
A^{1/2} (\xv^*, \xv_1^*, dt ,\lambda^*)  
  &=&
  (- dx^i - dt ( F^i _{(I \! R)}(\yv, \lambda)  - F^i _{(R)}(\yv, \lambda)  
 -  \partial_k  B^{ik}(\yv, \lambda ) ) ) 
  \nonumber\\
\times  \frac{ B^{-1}_{ij}  (\yv, \lambda) }{4 \,dt} 
 &\times&  ( - dx^j - dt ( F^j_{(I \! R)}(\yv, \lambda)  - F^j_{(R)}(\yv, \lambda)  
  -  \partial_l   B^{jl}(\yv, \lambda)  ) )
\nonumber\\
&+& \frac{1}{2} dt \, \partial_i ( F^i _{(I \! R)}(\yv, \lambda)  - F^i _{(R)}(\yv, \lambda)  )
 -  \frac{1}{4}  dt \, \partial_i \partial_j  B^{ij}(\yv, \lambda) .  
 \label{action-back-2}
\ea 

{Now dividing Eq.~(\ref{tran-prob-3-app}) by Eq.~(\ref{tran-prob-back}), taking the logarithm, and using the following symmetry properties, which follow immediately from Eqs.~(\ref{g-v-reversal-invariance}) and (\ref{DB-condition-2-B}):}
\ba
dv (\xv_1 ) &=&
 \sqrt{g(\xv_1)} d\mu(\xv_1), \\
 dv (\xv^*) &=& \sqrt{g(\xv^*)} d^n \xv^*
  = \sqrt{g(\xv)} d^n \xv, \\
  d\mu( \xv) &=& d\mu( \xv^*), \\
 \det B^{ij}(\yv^*, \lambda^*) &=&
 \det B^{ij}(\yv, \lambda),
\ea 
we obtain 
\ba
\log \frac{\pdf_F (\xv_1 , t + dt | \xv ,t) }
{\pdf_B(\xv^* , - t  | \xv_1^* , - t - dt)}
\cdot \frac{\sqrt{g(\xv_1)}} {\sqrt{g(\xv )}}
=  A^{1/2} (\xv^*, \xv_1^*, dt ,\lambda^*)  
-  A^{1/2} (\xv_1, \xv, dt,\lambda). 
\label{log-p-relation}
\ea
The RHS can be calculated readily using Eqs.~(\ref{action-forward-1}) and (\ref{action-back-2}):
\ba
\mbox{RHS of Eq.~(\ref{log-p-relation})} 
=  (dx^i - dt \, F^i _{(R)}(\yv, \lambda)) B_{ij}^{-1}(\yv, \lambda) 
( F^j_{(I \! R)}(\yv, \lambda) - \partial_l B^{jl} (\yv, \lambda) )
- dt \, \partial_i F^i _{(R)}(\yv, \lambda).  
\ea
Further using Eqs.~(\ref{F-decomp-def-3}) and (\ref{F-decomp-def-4}), we obtain  
\ba
&& A^{1/2} (\xv^*, \xv_1^*, dt ,\lambda^*)  
-   A^{1/2} (\xv_1, \xv, dt,\lambda)
= -  dx^i \, \partial_i \!\left[  \U(\yv, \lambda) - \log \sqrt{g(\yv)} \right] . 
\label{blahblah}
\ea 
{Note that $\yv = \xv_{1/2} = \xv + d \xv/2$.}  For any smooth function  $\Psi(\xv)$,  we have:
\be
dx^i \, \partial_i \Psi(\xv_{1/2}) 
= \Psi(\xv + d \xv) - \Psi(\xv) + O(d\xv^3).  
\ee
{We do not need to worry about $O(d\xv^3)$ terms, since they scale as $dt^{3/2}$ and do not contribute to the dynamics in the continuum limit.  
Hence the RHS of Eq.~(\ref{blahblah}) can be written as 
\be
- \U(\xv  +  d \xv , \lambda) + \U(\xv, \lambda) 
+ \log \sqrt{g(\xv  +  d \xv )} 
- \log \sqrt{g(\xv)}
= - d_{\xv} \U(\xv, \lambda) +  d_{\xv} \log \sqrt{g(\xv)} ,
\ee
where $d_{\xv}$ is defined in Eqs.~(\ref{dW-dQ-path-def-1}).  } Substituting these results back into Eq.~(\ref{log-p-relation}) we finally obtain 
\ba
\log \frac{\pdf_F (\xv_1 , t + dt | \xv ,t) }
{ \pdf_B (\xv^* , - t  | \xv_1^* , -t - dt)}
=  - d_{\xv} \U(\xv, \lambda) = - \beta \dbar {\mathcal Q},  
\label{EPF-app}
\ea
which is precisely Eq.~(\ref{DB-condition-3-new}).   {As explained in Sec.~(\ref{sec:Heat Formula}), the time-dependence of $\lambda$ needs not to be worried, since it only contributes only to the order $dt^{3/2}$.  }



\end{widetext}

\end{document}